%% file: main.tex
\documentclass[12pt]{article}
\usepackage{jheppub} % for details on the use of the package, please
\usepackage{graphicx}
\usepackage{amsmath}
\usepackage{bm}
\usepackage{amssymb}
\usepackage{slashed}
\usepackage{comment}
\usepackage{graphicx}
\usepackage{physics}
\usepackage{cancel}
\usepackage{comment}
\usepackage{mathtools}
\usepackage{empheq}
\usepackage[dvipsnames]{xcolor}
\usepackage{amsmath}
\usepackage{feynmp-auto}
\usepackage{tikz}
\usetikzlibrary{shapes.arrows}
\usepackage{feynarts}

\definecolor{darkred}{rgb}{0.5,0.0,0.0}
\definecolor{darkblue}{rgb}{0.0,0.0,0.9}
\definecolor{darkerblue}{rgb}{0.0,0.0,0.5}
\definecolor{darkgreen}{rgb}{0.0,0.5,0.0}
\definecolor{black}{rgb}{0.0,0.0,0.0}
\definecolor{brown}{rgb}{0.6,0.4,0.2}

\usepackage[colorlinks=true]{hyperref}
\hypersetup{
    linktocpage,
     colorlinks,
     citecolor=darkgreen,
     linkcolor= darkgreen,
     urlcolor=darkgreen
}
\allowdisplaybreaks

\newcommand{\dlipsred}[2]{\frac{d^{#2}#1}{(2\pi)^{#2}}}
\newcommand{\vp}[1]{\vec{#1}_{\perp}}

\newcommand{\nureg}[1]{\left(\frac{\nu}{#1}\right)^\eta}
\newcommand{\nb}{\bar{n}}

\newcommand{\logq}{\ln\frac{\mu^2}{-t}}
\newcommand{\logp}{\ln\frac{\nu}{Q}}
\newcommand{\eps}{\varepsilon}
\newcommand{\epsuv}{\eps_\text{UV}}
\newcommand{\epsir}{\eps_\text{IR}}

\newcommand{\ppsl}{\slashed{p}_\perp}
\newcommand{\cdt}{\!\cdot\!}
\def\cO{\mathcal{O}}
\def\cI{\mathcal{I}}
\def\cM{\mathcal{M}}

\def\cP{\mathcal{P}}
\def\cB{\mathcal{B}}

\title{Quark-Gluon Backscattering in the Regge Limit at One-Loop}

\author[a]{Arindam Bhattacharya,}
\author[b]{Aneesh V.~Manohar,}
\author[a]{and Matthew D.~Schwartz} 

\affiliation[a]{Department of Physics, Harvard University, Cambridge, MA 02138}

\affiliation[b]{Department of Physics, University of California San Diego, 9500 Gilman Drive, La Jolla, CA 92093-0319}

\emailAdd{arindamb@g.harvard.edu}
\emailAdd{schwartz@g.harvard.edu}
\emailAdd{amanohar@ucsd.edu}

\abstract{
At small momentum transfer, the quark-gluon scattering cross section ${\rm d}\sigma/{\rm d}t$ has a power-law divergence in the backward scattering region where the outgoing quark is nearly collinear to the incoming gluon. In this Regge limit $|t| \ll s$, the leading behavior of the $2\to 2$ amplitude can be described by the exchange of Glauber quarks. In Soft-Collinear Effective Theory (SCET) at leading power, Glauber quark exchange is given by five non-local Glauber quark operators, of which only one is generated at tree-level. We show that at leading power the QCD amplitude for quark-gluon backscattering at one-loop can be exactly reproduced by SCET using the tree-level Glauber operator.
The agreement between QCD and SCET of the ultraviolet, infrared, and rapidity divergences as well as all logarithms, Glauber phases and finite parts for all polarizations of the external gluons is a strong check on the effective theory. We find that the entire one-loop matching vanishes --- there is no correction to the operator generated at tree-level, and the coefficients of the other four operators remain zero at one-loop. This suggests that SCET with Glauber operators may be useful for uncovering new aspects of Regge physics in a systematically improveable way.
}

\begin{document}

\maketitle

\begin{fmffile}{feyngraph}
\unitlength = 1mm

%\unitlength = 0.4mm

%\thispagestyle{empty}

%\maketitle

\input{Introduction}

\input{GlauberSCET}

\input{rapidityreg}

\input{tree}

\input{oneloop}

\input{soft}

\input{boxes}

\input{zerobin}

\input{matching}

\input{conclusions}

\newpage
\appendix
\input{Wilson}
\input{feynmanrules}
\input{zerobintoy}
\input{integrals}

\end{fmffile}
	
\bibliographystyle{JHEP}
\bibliography{ref}

\end{document}

%% file: Introduction.tex
\section{Introduction}

The scattering of quarks and gluons at small momentum transfer is one of the most challenging processes to completely understand. In this kinematical region, called the Regge limit, two incoming quarks or gluons scatter into two outgoing jets moving in nearly the same direction as the incoming partons. The power-law behavior of the scattering amplitude, typically of the form ${s}/{t}$, makes the Regge limit dominate the cross sections. The ${s}/{t}$ behavior is present already at tree-level; at higher orders in perturbation theory it is supplemented by logarithms of the form $\ln({s}/{t})$. When the logarithms are resummed, the amplitude scales like $({s}/{t})^\gamma$ where $\gamma$ is called the Regge trajectory~\cite{Gell-Mann:1964aya,Mandelstam:1965zz,McCoy:1976ff,Grisaru:1973wbb,Fadin:1975cb,Lipatov:1976zz,Lipatov:1985uk}. The structure of scattering amplitudes in the Regge limit is of interest in a wide variety of applications, including the total cross sections at colliders~\cite{Pancheri:tot}, unitarity constraints, integrability~\cite{Lipatov:1993yb,Faddeev:1994zg}, factorization-violation~\cite{DelDuca:2001gu,DelDuca:2013ara,DelDuca:2014cya,DelDuca:2011ae}, and bootstrapping scattering amplitudes in ${\mathcal N}=4$ super-Yang Mills theory~\cite{Caron-Huot:2019vjl,DelDuca:2018rhj,DelDuca:2016lad,Basso:2014pla,Dixon:2011pw,Bartels:2014mka}. 

To study the Regge limit, the traditional approach has been essentially diagrammatic. One considers all the possible Feynman diagrams that could contribute, expands them in the Regge limit in some particular gauge, and looks for patterns and relationships among the contributions. This method allows one to extract the all-orders structure of the Regge limit and leads to the Balitsky–Fadin–Kuraev–Lipatov (BFKL) equation~\cite{Kuraev:1976ge,Balitsky:1978ic}. A number of textbooks describe this procedure in detail~\cite{kovchegov,forshaw}.

An alternative approach to studying the Regge limit is with effective field theory. Soft-Collinear Effective Theory (SCET)~\cite{bauerBXs,bauer2001effective,bauer2001invariant,bauer2002hard,bauer2002power,bauer2002soft} is a powerful tool which has been shown to reproduce the leading infrared singular behavior of QCD away from the Regge region. If none of the outgoing particles are collinear to any of the incoming particles, then scattering processes factorize into a number collinear sectors, one associated with each incoming parton or outgoing jet, plus a single global soft sector. It has been proven that all of the infrared divergences of QCD are reproduced by SCET in such configurations, and for infrared-safe cross sections, the leading power behavior is entirely reproduced. Applications of SCET to collider physics are diverse and in a number of cases resummation at the next-to-next-to-next-to-leading logarithmic level (N3LL) is possible. 

In the Regge limit, soft and collinear modes are insufficient to describe the leading power behavior of scattering amplitudes. Instead, one must supplement the SCET Lagrangian with a set of non-local potential operators, heuristically of the form $\cO_n \cO_s \cO_{\nb}/t$, 
where $\cO_n$ and $\cO_{\nb}$ encode the collinear scattering in the $n$ and $\nb$ directions, and $\cO_s$ encodes the soft scattering. The explicit ${1}/{t}$ dependence comes from integrating out the tree-level Glauber exchange~\cite{RS}. We refer to this theory as Glauber-SCET. Unlike in the hard-scattering case, where the Regge region is avoided, there are no rigorous proofs that SCET supplemented with Glauber potential operators of this type will indeed reproduce the leading power behavior of full QCD. Thus it is important to perform as many checks as possible. Calculations with Glauber-SCET are significantly more challenging than those in SCET without Glauber modes due to the presence of rapidity divergences. 

In the forward scattering region for quark-gluon scattering $qg \to qg$ with the outgoing quark collinear to the incoming quark, the Regge region is dominated by the exchange of Glauber gluons. These gluons have spacelike momentum $q$ with $0 \le -q^2 = -q_\perp^2 \ll Q^2$, where $Q$ is the center-of-mass energy of the collision. Including spin effects, the $qg \to qg$ forward scattering amplitude in the Glauber region scales like $({-q_\perp^2}/{Q^2})^{-1}$~\cite{RS}.\footnote{Here we are using the scaling for the spinors and not for the quark fields. The spinors power count as $\cO(({-q_\perp^2}/{Q^2})^0)$, while the quark field power counts as $\cO(({-q_\perp^2}/{Q^2})^\frac{1}{2})$.} In this paper, we instead focus on the backward scattering region. In this region, the Regge limit is described by Glauber fermion exchange and the amplitude scales like $({-q_\perp^2}/{Q^2})^{-1/2}$. Thus Glauber fermion exchange is power suppressed compared to Glauber gluon exchange. However, the two exchanges contribute in different kinematical regions; Glauber fermion exchange describes the leading power behavior in the backward scattering region. Moreover, Glauber gluon exchange is necessarily non-Abelian whereas Glauber-fermion exchange is not: in QED, the amplitude for $e^- \gamma \to e^- \gamma$ is nonsingular in the forward scattering region, but is singular for backward-scattering.
%, leading to logarithms $\ln (Q/m_e)$ of the electron mass $m_e$ in the total Compton cross section.  
Thus to make progress on some challenging foundational questions in quantum field theory, such as how to define the $S$-matrix for charged particles~\cite{Frye:2018xjj,Hannesdottir:2019opa,Hannesdottir:2019rqq}, a problem even in QED, it helps to have a better understanding of singularities present in QED, such as in the backward scattering region.

In this paper, we focus on quark-gluon backscattering in the Regge limit at one-loop order. We compute the full one-loop $2\to 2$ scattering amplitude in QCD and then expand in the Regge limit. We also compute the full one-loop $2 \to 2$ scattering amplitude in Glauber-SCET including only the leading order operator required for tree-level matching.
We find that the two calculations agree exactly.
% : all the ultraviolet (UV) divergences, infrared (IR) divergences and rapidity divergences agree (QCD is rapidity finite). 
% We find that the two calculations agree almost exactly: all the ultraviolet (UV) divergences, infrared (IR) divergences and rapidity divergences agree (QCD is rapidity finite). The only difference is that the Glauber-quark operator must get a matching correction proportional to $\alpha_s C_F$, which is also present for QED. No such correction was needed for quark-quark forward scattering through Glauber gluon exchange~\cite{RS}. In the conclusions we comment on the importance of this result and its implications.

%% file: GlauberSCET.tex
\section{Glauber SCET}
This section reviews the general construction of SCET, and describes the additional terms that need to be included in the Lagrangian to accomodate Glauber quark exchanges. It also serves to define the notation we use.

\subsection{SCET}
Soft-Collinear-Effective Theory is an effective theory that was developed to better understand the soft and collinear limit of gauge theories like QCD, systematize the concept of factorization, and successfully apply these ideas to collider physics. In order to describe soft and collinear modes, one sets up a light cone momentum basis, where the lightlike elements of the basis are taken to align with the jet axis in a collider process. For any lightlike direction $n^\mu=(1,\vec{n})$ one can define the backwards lightlike direction $\nb^\mu = (1,-\vec{n})$ and two transverse directions denoted by $\perp$ that are orthogonal to $n^\mu$ and $\nb^\mu$. Any four-vector $p^\mu$ can be decomposed into lightcone and perpendicular components,
\begin{align}
     p^\mu &= \frac{\nb \cdot p}{2}n^\mu +\frac{n \cdot p}{2}\nb^\mu +p_\perp^\mu \,,
\end{align}
where $n\cdt \nb=2$ and
\begin{align}
    n\cdot n&=\nb\cdot \nb=n\cdot p_\perp=\nb \cdot p_\perp=0 \, .
\end{align}
Sometimes the $p^\pm$ notation is also used for the lightcone components,
\begin{align}
    p^+  &= n\cdot p, & p^- &= \nb \cdot p\,.
\end{align}
%In the limit where the scattering is forward/backward, we have only two lightlike directions $n$ and $\nb$, which we align with the direction of the incoming quark (outgoing gluon), and incoming gluon (outgoing quark) respectively. 

SCET describes scattering process in an expansion in a power counting parameter $\lambda \ll 1$. For jet physics, the parameter might be the ratio of the mass of an outgoing jet to the center-of-mass energy, $\lambda \sim {m}/{Q}$. For Regge physics, the power counting parameter is the transverse momentum of the scattered particles (relative to the incident direction), divided by the center-of-mass energy, $\lambda^2 \sim {\vec{q}_\perp^2}/{Q^2}$.  Modes in the effective theory are defined by the scaling of their lightcone momentum components with respect to the power counting parameter. Using the notation
\begin{equation}
    p^\mu \sim ( n\cdot p,\, \nb \cdot p,\, |p_\perp|) \equiv (p^+, p^-, |p_\perp|)\,,
\end{equation}
the various soft and collinear modes are
\begin{equation}
\begin{aligned}
n\ \text {collinear}:\ p^{\mu} & \sim Q\left(1, \lambda^{2}, \lambda\right) \\
\bar{n}\ \text {collinear}:\ p^{\mu} & \sim Q\left(\lambda^{2}, 1, \lambda\right) \\
\text{soft}:\ p^{\mu} & \sim Q(\lambda, \lambda, \lambda)\\
\text{ultrasoft}:\ p^{\mu} & \sim Q(\lambda^2, \lambda^2, \lambda^2)\,.
\end{aligned}
\end{equation}
Whether soft or ultrasoft modes are relevant depends on the particular problem one is interested in. Ultrasoft modes only affect the smallest lightcone component of collinear momenta, but not the transverse components.  For most hard-scattering jet physics processes, ultrasoft modes are sufficient. SCET with ultrasoft but not soft modes is called $\text{SCET}_{\text{I}}$.  For Regge physics, we will need to account for recoil of the transverse components of collinear fields from soft emissions. For such physics, soft rather than ultrasoft modes are relevant. SCET with soft modes is called $\text{SCET}_{\text{II}}$. The invariant mass of collinear modes is $\lambda^2$, of soft modes is $\lambda^2$, and of ultrasoft modes is $\lambda^4$. The invariant mass of ultrasoft modes in $\text{SCET}_{\text{I}}$ is much smaller than that of collinear modes, whereas soft and collinear modes in $\text{SCET}_{\text{II}}$ have the same invariant mass.

In addition to the scaling of the momentum components, fields and states in the theory also have scaling which can depend on spin. A Dirac spinors $u(p)$ can always be decomposed into two lightcone spinors $u=\xi_n + \varphi_n$
using the projectors $\slashed{\nb}\slashed{n}/4$ and
$\slashed{n}\slashed{\nb}/4$,
\begin{align}
    \frac{\slashed{n} \slashed{\nb}}{4} u &= \xi_n,
    &    
    \frac{\slashed{\nb} \slashed{n}}{4} u &= \varphi_n\,.
\end{align}
Due to the equations of motion $\slashed{p} u(p)=0$, the $\varphi_n$ spinor is power suppressed compared to the $\xi_n$ spinor, $\varphi_n \sim \lambda$ and $\xi_n \sim \lambda^0$~\footnote{Note that the collinear quark fields, $\xi_{n}(x)$ and $\varphi_n(x)$ on the other hand scale as $\lambda$ and $\lambda^2$ respectively.}. Thus SCET integrates out $\varphi_n$ using the equations of motion, so that fermions are described by the spinor $\xi_n$, which has two independent components. This amounts to replacing
\begin{align}
    u &\to \left( 1 + \frac{\ppsl \slashed{\nb}}{4 \nb \cdot p} \right) \xi_n,
    &
    \bar{u} &\to \bar{\xi}_{\nb} \left( 1 + \frac{\slashed{n} \ppsl}{4 \nb \cdot p} \right) \,.
\end{align}
Similarly, gluon polarization states have scaling determined by consistency with collinear gauge invariance. The components of the polarization scale proportionally to the momenta,
\begin{align}
    n \cdot \epsilon &\sim \lambda, &
    \epsilon_\perp &\sim \lambda^0, &
    \nb\cdot \epsilon \sim \frac{1}{\lambda},
\end{align}
for an $n$-collinear gluon.
Using the equations of motion $p \cdot \epsilon = 0$ for an on-shell gluon, the small polarization component can be integrated out analogously to the fermion case,
\begin{equation}
    n\cdot \epsilon \to -\frac{n\cdot p }{\nb\cdot p}\, \nb\cdot \epsilon -  \frac{2}{\nb\cdot p}\, p_\perp \cdot \epsilon_\perp\,.
\end{equation}
Since the equations of motion are used to simplify the SCET Lagrangian (or equivalently, the SCET and QCD amplitudes only agree on-shell), we must consistently use these substitutions to see agreement between QCD and SCET. That is, we must eliminate the small components of the spinors and polarization vectors in QCD to extract the leading power behavior.

Operators in SCET are usually written in terms of collinear gauge invariant building block, constructed from the leading power spin components of the fields wrapped in Wilson lines. For quarks and gluons these combinations are~\cite{marcantonini2009reparametrization}\footnote{We use the sign convention $D_\mu = \partial_\mu - i g A_\mu$.}
\begin{align}
    \chi_{n}(x) &=\left[W_n^{\dagger}(x) \xi_{n}(x)\right], & \cB_{n \perp}^{\mu}(x) &=\frac{1}{g}\left[W_n^{\dagger}(x) i D_{\perp}^{\mu} W_{n}(x)\right],
    \label{2.10}
\end{align}
Here, $W_n(x)$ denotes a collinear Wilson line, representing the source of radiation from all charged particles other than the collinear field in the $n$ direction. Similarly, the gauge invariant soft gluon field is given by 
\begin{equation}
 %   \mathcal{B}_{S \perp}^{\bar{n} \mu}=\frac{1}{g}\left[S_{\bar{n}}^{\dagger} i D_{S \perp}^{\mu} S_{\bar{n}}\right], \quad 
 \mathcal{B}_{S \perp}^{n \mu}=\frac{1}{g}\left[S_{n}^{\dagger} i D_{S \perp}^{\mu} S_{n}\right]. \label{Sout}
\end{equation}
where $S_{n}$ denotes a soft Wilson line generated by the soft gluon fields. Analogous definitions hold for $\nb$-collnear and soft fields. 
Explicit forms of the Wilson lines and their origin is discussed in Appendix~\ref{sec:Wilson}. 

Equation~\eqref{Sout} are gluon fields for soft outgoing radiation.
One could alternatively define the gauge-invariant soft gluon fields in terms of Wilson lines representing incoming radiation:
\begin{equation}
%    \overline{\mathcal{B}}_{S \perp}^{\bar{n} \mu}=\frac{1}{g}\left[\overline{S}_{\bar{n}}^{\dagger} i D_{S \perp}^{\mu} \overline{S}_{\bar{n}}\right], \quad 
    \overline{\mathcal{B}}_{S \perp}^{n \mu}=\frac{1}{g}\left[\overline{S}_{n}^{\dagger} i D_{S \perp}^{\mu} \overline{S}_{n}\right].
\end{equation}
$\overline{S}_n$ differs from $S_n$ in whether the Wilson line is from $-\infty$ to $x$ or $x$ to $\infty$, i.e.\  the $\pm i0^+$ prescription in the eikonal propagators (see Appendix~\ref{sec:Wilson}). The quark-gluon scattering amplitude in our Glauber-SCET problem is independent of which Wilson lines are chosen. We choose outgoing Wilson lines as in Eq.~\eqref{Sout}. If one uses incoming Wilson lines, the soft graphs come out differently, but there is a non-vanishing soft-Glauber zero-bin subtraction~\cite{Manohar:2006nz} which restores equivalence to the outgoing case. This issue is discussed in~\cite{RS} and we reproduce their observation in Sections~\ref{sec:zerobin} and Appendix~\ref{app:zerobintoy}.

\subsection{Glauber operators}

In the Regge limit, the exchanged Glauber gluon or quark couples different collinear sectors to each other. For a Glauber particle to couple to the $n$ collinear sector, it must have subdominant scaling to the $n$-collinear scaling, i.e. $p_G \lesssim Q(1,\lambda^2,\lambda)$.  For the same Glauber particle to couple to the $\nb$-colliner sector, it must have $p_G \lesssim Q(\lambda^2,1,\lambda)$. Thus we have
\begin{equation}
    \text{Glauber}: p_G^\mu \sim Q (\lambda^2, \lambda^2, \lambda)
\end{equation}
for the momentum scaling of Glauber particles.
Since $p^2 = (\nb \cdot p)(n \cdot p) +p_\perp^2$ and the two terms scale like $\lambda^4$ and $\lambda^2$ respectively, it is impossible for a Glauber particle to be on-shell. Thus Glauber exchange in SCET is described by non-propagating potentials. Glauber potential operators were introduced in Ref.~\cite{RS} for Glauber gluon exchange, and Ref.~\cite{MSSV} for Glauber fermion exchange.

In the fermionic Glauber case, the operator required for tree-level matching is~\cite{MSSV}
\begin{equation}
    \cO_{T}=
%    \cFG \,
    \bar{\chi}_{n}\slashed{\cB}_{n\perp} \frac{1}{\slashed{\cP}_\perp} \cO_{s} \frac{1}{\slashed{\cP}_\perp} \slashed{\cB}_{\nb \perp} \chi_{\nb}\,, \label{OT}
\end{equation}
%with $\cFG =1$ \mds{why is this here?},
where $\cP^\mu_\perp$ is a label operator that picks out the transverse momenta of the fields it acts on. 
The operator $\cO_s$ involves soft fields:
\begin{equation}
    \cO_{s}=-2 \pi \alpha_{s}\left[S_{\bar{n}}^{\dagger} S_{n} \slashed{\cP}_{\perp}+\slashed{\cP}_{\perp} S_{\bar{n}}^{\dagger} S_{n}-S_{\bar{n}}^{\dagger} S_{n} g \slashed{\cB}_{S \perp}^{n}-g \slashed{\cB}_{S \perp}^{\bar{n}} S_{\bar{n}}^{\dagger} S_{n}\right]\,. \label{Os}
\end{equation}
The coefficients of the various terms in  this operator were derived in~\cite{MSSV} by matching onto QCD in the Regge limit with additional soft gluons emitted.

To describe soft-collinear scattering in the Regge limit,
soft-collinear Glauber operators are necessary~\cite{RS,MSSV}:
\begin{align}
    \cO_{Sn}&=(-4\pi\alpha_s)\ \bar{\chi}_n \slashed{\cB}_{n\perp}\frac{1}{\slashed{\cP}_\perp}\slashed{\cB}_{S\perp}^n\psi_{S}^n\ +\ \text{h.c.}, &    
    \cO_{S\nb} &=(-4\pi\alpha_s)\ \bar{\chi}_{\nb} \slashed{\cB}_{\nb\perp}\frac{1}{\slashed{\cP}_\perp}\slashed{\cB}_{S\perp}^{\nb}\psi_{S}^{\nb}\ +\ \text{h.c.},
    \label{eq:softglauber}
\end{align}
where
\begin{align}
    \psi_S^n&=S_n^\dagger \psi_S\ , &  \psi_S^{\bar{n}}&=S_{\bar{n}}^\dagger \psi_S\ .
\end{align}
Here $\psi_S$ is a soft quark field, and the label ($n$/$\nb$) denotes the direction of the Wilson line that dresses the field. 
As with $\cO_T$ and $\cO_s$ the structure of these operators has been fixed to reproduce the leading-power behavior of the full QCD amplitudes at tree level.
As shown in~\cite{MSSV} at one-loop, the interference between emission from these two operators in a time-ordered product
contributes to collinear-collinear scattering at one-loop (see Eq.~\eqref{Incoll} below).

The leading-order collinear-collinear Glauber gluon operator is
\begin{equation}
    \cO_{GG0}=\left(\bar{\chi}_{n}T^b \frac{\slashed{\bar{n}}}{2} \chi_n\right)\frac{1}{\cP_\perp^2} \cO^{bc}_{s}\frac{1}{\cP_\perp^2}\left(\bar{\chi}_{\nb}T^c \frac{\slashed{n}}{2} \chi_{\nb}\right),
\end{equation}
where $\cO^{bc}_{s}$ arises from a Glauber gluon exchange and can be found in~\cite{RS}. 
The structure and normalization of this operator is fixed at tree-level from matching to quark-quark scattering in the forward limit in QCD. As with the soft-collinear operator, this operator will contribute to quark-gluon backscattering at one loop. In this case, it is through interference with the collinear-collinear Glauber quark operator in box diagrams (see Eq.~\eqref{Iboxes} below).

In addition to the above operators, all of which were discussed in~\cite{MSSV}, there are additional leading-power collinear-collinear Glauber operators that could be relevant beyond tree-level. We can write the full Fermionic Glauber content of the Glauber-SCET Lagrangian as
\begin{equation}
    \cO_{FG} = C_T \cO_T + C_{L1} \cO_{L1}+ C_{L2} \cO_{L2}+ C_{L3} \cO_{L3}+ C_{L4} \cO_{L4} \label{OFG}
\end{equation}
where
\begin{align}
    \cO_{L 1} &=\bar{\chi}_{n} \slashed{\cB}_{n \perp} \frac{1}{\slashed{\cP}_\perp} \cO_{s} \frac{1}{\cP_{\perp}^{2}}
    \chi_{\nb}\left(\cP_{\perp} \cdot \cB_{\nb \perp}\right) \label{OL1}\,, \\
\cO_{L 2} &=\bar{\chi}_{n}\left(\cP_{\perp} \cdot \cB_{n \perp}\right) \frac{1}{\cP_{\perp}^{2}} \cO_{s} \frac{1}{\slashed{\cP}_\perp}\slashed{\cB}_{\nb \perp} \chi_{\nb} \,, \\
\cO_{L 3} &=\bar{\chi}_{n}\left(\cP_{\perp} \cdot \cB_{n \perp}\right)\  \frac{1}{\cP_{\perp}^{2}} \cO_{s} \frac{1}{\cP_{\perp}^{2}}\ \left(\cP_{\perp} \cdot \cB_{\nb \perp}\right) \chi_{\nb} \,, \\
\cO_{L 4} &=\bar{\chi}_{n} \cB_{n \perp}^{\mu}\  \frac{1}{\slashed{\cP}_\perp} \cO_{s} \frac{1}{\slashed{\cP}_\perp}\  \cB_{\nb \perp\mu} \chi_{\nb} \,. \label{OL4}
\end{align}
The soft operators appearing with these collinear-collinear Glauber operators could have different linear combinations of the operators in~\eqref{Os}. In general~\cite{MSSV}
\begin{multline}
\mathcal{O}_{s}=-4 \pi \alpha_{s}\left[\frac{C_{S1}}{2}\left(g \slashed{\cB}_{\perp s}^{n} S_{n}^{\dagger} S_{\bar{n}}+S_{n}^{\dagger} S_{\bar{n}} g \slashed{\cB}_{\perp s}^{\bar{n}}\right)+\frac{C_{S2}}{2}\left(S_{\bar{n}}^{\dagger} S_{n} g \slashed{\cB}_{S \perp}^{n}+g \slashed{\cB}_{S \perp}^{\bar{n}} S_{\bar{n}}^{\dagger} S_{n}\right)\right.\\*
\left.+\frac{C_{S3}}{2}\left(S_{n}^{\dagger} S_{\bar{n}} \slashed{\cP}_{\perp}+\slashed{\cP}_{\perp} S_{n}^{\dagger} S_{\bar{n}}\right)+\frac{C_{S4}}{2}\left(S_{\bar{n}}^{\dagger} S_{n} \slashed{\cP}_{\perp}+\slashed{\cP}_{\perp} S_{\bar{n}}^{\dagger} S_{n}\right)\right]
\label{Osgeneral}
\end{multline}
At tree-level $C_{S1}=0, C_{S2} =-1, C_{S3} = 0$ and $C_{S4} = 1$ so that Eq.~\eqref{Osgeneral} reduces to Eq.~\eqref{Os}.
Since quark-gluon backscattering at one-loop is not sensitive to the higher order structures in $\cO_s$, we write $\cO_s$ for all the soft operators within the $\cO_{Lj}$ for simplicity. Similarly, additional soft-collinear Glauber operators could be present, but they do not contribute to one-loop quark-gluon backscattering. Thus, at one-loop, there are 5 possible operators that could receive corrections, as indicated in Eq.~\eqref{OFG}.

%% file: rapidityreg.tex
\subsection{Rapidity Regulator}\label{etaregsec}

One distinguishing feature of SCET$_\text{II}$ is that all modes in the effective theory sit on the same mass (or virtuality) hyperbola and are distinguished by their rapidities alone. This is in contrast to SCET$_\text{I}$, where the ultrasoft modes have a distinct virtuality compared to the collinear modes. Thus, in SCET$_\text{II}$, one could in principle boost a collinear mode into a soft mode and vice-versa. This indistinguishability in terms of boosts shows up in divergences at the loop level in the EFT that are not regulated by regulators such as dimensional regularization. Such divergences are called rapidity divergences~\cite{chiu2012formalism}, and are a result of demanding strict factorization into modes that live on the same mass hyperbola. 

Rapidity divergences can be regulated, and the regulator must drop of out physical quantities, just like with ultraviolet or infrared divergences. In this paper, we use the $\eta$ regulator introduced in~\cite{chiu2012formalism}. 
Alternative regulator choices include the exponential regulator~\cite{duffexponentialreg}, and the delta regulator~\cite{chiudeltareg,Chiu:2009mg}. All such regulators explicitly break boost invariance (RPI-III symmetry) between the different modes in the EFT.

The $\eta$-regulator can be systematically included by modifying the soft and collinear Wilson lines. This is easiest to do in momentum space, where the rapidity-regulated outgoing soft and collinear Wilson lines take the form~\cite{ChiuRRGE,RS}
\begin{align}
S_{n} &=\sum_{\text {perms }} \exp \left(-\frac{g}{n \cdot \mathcal{P}} \frac{\left|2 \mathcal{P}^{z}\right|^{-\eta / 2}}{\nu^{-\eta / 2}} n \cdot A_{s}\right)\,, \\
W_{n} &=\sum_{\text {perms }} \exp \left(-\frac{g}{\bar{n} \cdot \mathcal{P}} \frac{|\bar{n} \cdot \mathcal{P}|^{-\eta}}{\nu^{-\eta}} \bar{n} \cdot A_{n}\right)\,.
\end{align}
Here $\nu$ is a new scale that is analogous to $\mu$ in dimensional regularization. It must drop out of physical quantities at fixed order. A given sector has only a single rapidity scale, so one can choose $\nu$ equal to that scale to remove the logarithms in each sector separately. However, different sectors can have different rapidity scales. Then the rapidity renormalization group can be used to resum the rapidity logarithms by evolving all sectors to their individual scales, starting from a common rapidity scale. Systematization of such rapidity RGEs was first done in~\cite{chiu2012formalism,ChiuRRGE}. Later work (Ref.~\cite{RS} and then Ref.~\cite{MSSV}) formulated the BFKL equations as rapidity RGEs in the language of SCET. We will not attempt to resum any large logarithms in this paper.

Care needs to taken while using the $\eta$-regulator. Since one encounters rapidity divergences, UV divergences and IR divergences, a prescription for order of limits is required. In order to remain on the mass hyperbola while sending the rapidity cutoff to it limit. The correct order of limits is to first take $\eta \rightarrow 0$, so that ${\eta}/{\epsilon^n}\rightarrow 0$ for any $n$, and only then take the $\epsilon \rightarrow 0$ limit of dimensional regularization~\cite{chiu2012formalism}.

The sum of the soft and collinear contributions at fixed order are $\eta$ and $\nu$ independent. This follows from the observation that rapidity divergences  arise from a strict delineation of modes in SCET. If SCET is to reproduce the full QCD amplitude at leading power, which is rapidity finite, it must therefore be rapidity finite as well.

%% file: tree.tex
\section{Tree-level matching \label{sec:tree}}
We begin with matching at tree-level between SCET and QCD. That is, we demonstrate that the entire QCD amplitude is reproduced by SCET at leading power in the Regge limit. This will establish the procedure and notation we use for matching at one-loop in the next section. We follow the analysis in Ref.~\cite{MSSV}.

\subsection{Kinematics of backwards scattering}
We denote the incoming quark and gluon momenta as $p_1$ and $k_1$ and the outgoing quark and gluon momenta as $p_2$ and $k_2$, as indicated in Figure~\ref{fig:scatfig}.
\input{diagrams/scattfig}
We work in the centre of mass frame and split the momentum transfer evenly between the incoming and outgoing particles:
\begin{align}
     p_1^\mu &= \frac{Q}{2} n^\mu + \frac{1}{2} q_\perp^\mu + \frac{\vec{q}_\perp^2}{8 Q} \nb^\mu,
     & 
     k_1^\mu &= \frac{Q}{2} \nb^\mu - \frac{1}{2} q_\perp^\mu + \frac{\vec{q}_\perp^2}{8 Q} n^\mu \,, \nonumber \\
    p_2^\mu &= \frac{Q}{2} \nb^\mu + \frac{1}{2} q_\perp^\mu + \frac{\vec{q}_\perp^2}{8 Q} n^\mu,
     &
      k_2^\mu &= \frac{Q}{2} n^\mu - \frac{1}{2} q_\perp^\mu + \frac{\vec{q}_\perp^2}{8 Q} \nb^\mu \,.
      \label{3.1}
\end{align}
In this case,
\begin{equation}
    q^\mu = p_1^\mu - k_2^\mu = p_2^\mu - k_1^\mu = q_\perp^\mu\,.
\end{equation} 
An advantage of these coordinates is that the momentum transfer is purely transverse. In a different frame, where $p_{1\perp}=0$ for example, the Glauber momentum $q$ would have to have a nonzero $n\cdot q$ component to keeping the external particle momenta on-shell. The results for the matching are frame-independent, but the choice Eq.~\eqref{3.1} makes the calculation marginally simpler.
We also write
\begin{equation}
    q^2 = q_\perp^2 = -\vec{q\,}_\perp^2 = t < 0\,.
\end{equation}
We will use $q_\perp^2$ and $t$ interchangeably.

To integrate out the small spin and polarizations we use the equations of motion 
\begin{equation}
    \slashed{p}_1 u(p_1)=\slashed{p}_2 u(p_2)= k_1 \cdot \epsilon_1 = k_2 \cdot \epsilon_2 = 0 \,,
\end{equation}
which allows us to substitute
\begin{align}
    u(p_1) &= \left(1 + \frac{\slashed{q}_\perp \slashed{\nb} }{4 Q}\right) \xi_n, &
    \bar{u}(p_2) &= \bar{\xi}_{\nb} \left(1 + \frac{\slashed{n} \slashed{q}_\perp  }{4 Q}\right)\,, \label{ueom}\\
    \nb \cdot \epsilon_1 &= - \frac{\vec{q}_\perp\cdot \vec{\epsilon}_{1\perp}}{Q} - \frac{\vec{q\,}_\perp^2}{4 Q^2} n\cdot \epsilon_1, &
    n \cdot \epsilon_2 &= - \frac{\vec{q}_\perp\cdot \vec{\epsilon}_{2\perp}}{Q} - \frac{\vec{q\,}_\perp^2}{4 Q^2} n\cdot \epsilon_1 \,.\label{Aeom}
\end{align}
The scaling of the remaining objects that can appear in matrix elements is
\begin{align}
    Q &\sim 1, & 
    q_\perp &\sim \lambda,    &
    \xi_n &\sim \bar{\xi}_{\nb} \sim \lambda^0, &
    \epsilon_{1\perp} &\sim \epsilon_{2\perp} \sim \lambda^0, &
    n\cdot \epsilon_1 &\sim \nb \cdot \epsilon_2 \sim \lambda^{-1}\,.
\end{align}
Despite the anticipated ${1}/{t} \sim \lambda^{-2}$ kinematic behavior and the $\lambda^{-1}$ scaling of the nearly-forward polarized gluons,  the matrix elements for quark-gluon scattering in the Regge limit will scale at most like $\lambda^{-1}$.
The possible $\cO(\lambda^{-1})$ matrix elements with these ingredients are 

\begin{align}
    \cM_{0} &=- \xi_{\nb}\slashed{\epsilon}_{1\perp}\frac{\slashed{q}_\perp}{t}\slashed{\epsilon}_{2\perp}\xi_{n} \,, \\
    \cM_{1\nb}&= - \xi_{\nb}\slashed{\epsilon}_{1\perp}\xi_{n} \frac{\nb \cdt \epsilon_{2}}{Q}, &
    \cM_{2n} &=- \xi_{\nb}\slashed{\epsilon}_{2\perp}\xi_{n} \frac{n \cdt \epsilon_{1}}{Q} \,,\\
    \cM_{1} &=- \xi_{\nb}\slashed{\epsilon}_{1\perp}\xi_{n}\frac{q_\perp\cdt \epsilon_{2\perp}}{t}, & \cM_{2} &=-g^2\ \xi_{\nb}\slashed{\epsilon}_{2\perp}\xi_{n}\frac{q_\perp\cdt \epsilon_{1\perp}}{t}\,,\\
    \cM_{q1} &=- \xi_{\nb}\slashed{q}_{\perp}\xi_{n}\frac{(q_\perp\cdt \epsilon_{2\perp})n \cdt \epsilon_{1}}{tQ}, &
    \cM_{q2} &=- \xi_{\nb}\slashed{q}_{\perp}\xi_{n}\frac{(q_\perp\cdt \epsilon_{1\perp})\nb \cdt \epsilon_{2}}{tQ} \,, \\
    \cM_{q12} &=- \xi_{\nb}\slashed{q}_{\perp}\xi_{n} \frac{(n \cdt \epsilon_{1})(\nb \cdt \epsilon_{2})}{Q^2} \,, &
    \cM_{q12\perp} &=- \xi_{\nb}\slashed{q}_{\perp}\xi_{n} \frac{\epsilon_{1\perp}\cdt \epsilon_{2\perp}}{t},  \\
    \cM_{qqq} &=- \xi_{\nb}\slashed{q}_{\perp}\xi_{n} \frac{(q_\perp \cdt\epsilon_{1\perp})(q_\perp \cdt\epsilon_{2\perp})}{t^2}\,.
\end{align}
For the amplitude to satisfy the Ward identity, only certain linear combinations of these operators are allowed. For example, the tree-level matrix element of the gauge-invariant operator $\cO_{T}$ in Eq.~\eqref{OT} is
$g^2 \cM_{T}$,
\begin{align}
     \cM_{T} &\equiv\cM_{0}+\frac{1}{2}\left(\cM_{1\nb}+\cM_{2n}\right)+\frac{1}{4}\cM_{q12}\\*
     &=- \xi_{\nb}\left[\slashed{\epsilon}_{1\perp}-\slashed{k}_{1\perp}\frac{n \cdt \epsilon_{1}}{n \cdt k_1}\right]\frac{\slashed{q}_\perp}{t}\left[\slashed{\epsilon}_{2\perp}-\slashed{k}_{2\perp}\frac{\nb \cdt \epsilon_{2}}{\nb \cdt k_2}\right]\xi_{n}\,.
\end{align}
It is easy to see that upon replacing $\epsilon_1 \to k_1$ or $\epsilon_2 \to k_2$ this amplitude vanishes. The matrix element combinations corresponding to the other gauge invariant operators in Eqs.~\eqref{OL1} to Eq.~\eqref{OL4} are
\begin{align}\label{completeops}
    \cM_{L1} &\equiv \cM_{1}+\frac{1}{2}\left(\cM_{1\nb}+\cM_{q1}\right)+\frac{1}{4}\cM_{q12},\\
    \cM_{L2} &\equiv \cM_{2}+\frac{1}{2}\left(\cM_{2n}+\cM_{q2}\right)+\frac{1}{4}\cM_{q12},\\
    \cM_{L3}&\equiv\cM_{q12\perp}+\frac{1}{2}\left(\cM_{q1}+\cM_{q2}\right)+\frac{1}{4}\cM_{q12},\\
    \cM_{L4}&\equiv\cM_{qqq}+\frac{1}{2}\left(\cM_{q1}+\cM_{q2}\right)+\frac{1}{4}\cM_{q12}\,.
\end{align}
Note that if we consider only $\epsilon_\perp$ polarizations,  only the first term in any of these combinations contributes. However, it is important to check the matching for all polarizations as it elucidates additional aspects of the effective theory (such as the need for non-vanishing zero-bin subtractions) as we will see in the next section. 

\subsection{Tree-level graphs}

In QCD, at tree-level, three graphs contribute. The $t$-channel graph is ($a$ is the incoming gluon color and $b$ is the outgoing gluon color)
\begin{equation}
   i \cI_{a} = 
    \begin{gathered} \parbox{20mm} {\resizebox{20mm}{!}{
    \begin{fmfgraph*}(50,30)
	\fmfleft{L1,L2}
	\fmfright{R1,R2}
	\fmf{fermion}{L2,v1,v2,R1}
	\fmf{gluon}{L1,v2}
	\fmf{gluon}{v1,R2}
    \end{fmfgraph*}}}
    \end{gathered}
    = -i g^2 \frac{T^a T^b}{t}\bar{u}(p_2) \slashed{\epsilon_1}(\slashed{p_2} - \slashed{k_1}) \slashed{\epsilon_2} u(p_1)\,.
\end{equation}
To expand at leading power we apply Eq.~\eqref{ueom} and Eq.~\eqref{Aeom}, and keep only terms at leading power in $\lambda$.\footnote{At tree-level the small polarizations $n\cdot\epsilon_1$ and $\nb\cdot \epsilon_2$ do not contribute at leading power, so using Eq.~\eqref{Aeom} in this case is not needed.}
This leads to
\begin{equation}
     \cI_{a} \cong g^2 T^a T^b \left[\cM_T - \cM_{1\nb} - \cM_{2n}\right]\,.
\end{equation}
The $s$-channel graph gives
\begin{align}
      i \cI_{b} &=
   \begin{gathered} \parbox{20mm} {\resizebox{20mm}{!}{
    \begin{fmfgraph*}(50,30)
	\fmfleft{L1,L2}
	\fmfright{R1,R2}
	\fmf{fermion}{L2,v1,v2,R1}
	\fmf{gluon}{L1,v1}
	\fmf{gluon}{v2,R2}
    \end{fmfgraph*}}}
    \end{gathered}
     = -i g^2 \frac{T^b T^a}{Q^2} \Big[ 2 (p_2 \cdot \epsilon_2) \bar{u} \slashed{\epsilon}_1 u + \bar{u} \slashed{\epsilon}_2 \slashed{k}_2 \slashed{\epsilon}_1 u \Big]\,, \nonumber \\*
    &\cong g^2 T^b T^a \left[ \cM_{1\nb} + \cM_{2n} \right]\,.
\end{align}
The $u$-channel graph is
\begin{align}
    i \cI_{c} &= 
    \begin{gathered} \parbox{20mm} {\resizebox{20mm}{!}{
    \begin{fmfgraph*}(50,30)
	\fmfleft{L1,L2}
	\fmfright{R1,R2}
	\fmf{phantom}{L2,v1}
	\fmf{phantom,tension=0.5}{v1,v2}
	\fmf{phantom}{v2,R1}
	\fmf{phantom}{L1,v2}
	\fmf{phantom}{v1,R2}
	\fmffreeze
	\fmf{fermion}{L2,v1,R1}
	\fmf{gluon}{L1,v2,v1}
	\fmf{gluon}{v2,R2}
	\end{fmfgraph*}}}
    \end{gathered} \nonumber \\*
    & = i g^2 \frac{i f^{abc} T^b}{-Q^2-t} \Big[ (k_2 +p_1 - p_2)\cdt \epsilon_1 \bar{u} \slashed{\epsilon}_2 u-(k_1-p_1+p_2)\cdot \epsilon_2 \bar{u} \slashed{\epsilon}_1 u + \epsilon_1 \cdt \epsilon_2 (\slashed{k}_1 + \slashed{k}_2) u\Big] \,, \nonumber \\*
    &\cong i g^2 (T^a T^b -T^b T^a)\left[ \cM_{1\nb} + \cM_{2n} \right]\,,
\end{align}
so that the total tree-level amplitude is
\begin{equation}
   \cI_{\text{QCD}}^0= \cI_a + \cI_b + \cI_c = g^2 T^a T^b \cM_T \,.
\end{equation}
Note that all 3 diagrams contribute at leading power, although only the $t$-channel fermion exchange diagram contributes to the production of $\epsilon_\perp$.

At tree level, we have a single diagram in the EFT given by the matrix element of operator $\cO_{T}$ in Eq. \eqref{OT}. The contributions are
\begin{align}
&    \bra{\epsilon_2^b}\slashed{\cB}_{n\perp}\chi_n\ket{\xi_n} =
    \parbox{20mm} {\resizebox{20mm}{!}{
  \begin{fmfgraph*}(30,20)
    \fmfset{arrow_len}{3mm}
    \fmfset{arrow_ang}{20}
    \fmfstraight
    \fmftop{o1,d1,i1}
    \fmfbottom{i2,d2,o2}
    \fmflabel{$\epsilon_2^b(k_2)$}{i1}
    \fmflabel{$\xi_n(p_1)$}{o1}
    \fmf{phantom}{i2,v2}
    \fmf{phantom}{v2,o2}
    \fmf{fermion}{o1,v1}
    \fmf{gluon}{v1,i1}
    \fmffreeze
    \fmf{plain}{v1,i1}
    \fmf{dbl_dots_arrow,f=(0.8,,0.1,,0.1),label=$q$}{v1,v2}
    \fmfv{d.sh=circle, d.fi=full, d.si=2thick, f=(0.8,,0.1,,0.1)}{v1}
  \end{fmfgraph*}}}
    = T^b\left(\slashed{\epsilon}_{2\perp}-\slashed{k}_{2\perp} \frac{\nb\cdot \epsilon_2}{\nb\cdot k_2}\right)\xi_n \,,
\\
&          \bra{0}\cO_s\ket{0} 
          =
    \parbox{20mm} {\resizebox{20mm}{!}{
  \begin{fmfgraph*}(30,20)
    \fmfset{arrow_len}{3mm}
    \fmfset{arrow_ang}{20}
    \fmfstraight
    \fmfbottom{b1}
    \fmftop{t1}
    \fmf{dbl_dots_arrow,f=(0.8,,0.1,,0.1),label=$q$}{t1,v1,b1}
    \fmfv{d.sh=circle, d.fi=full, d.si=2thick, f=(0.8,,0.1,,0.1)}{v1}
  \end{fmfgraph*}}}
    =-ig^2 \slashed{q}_\perp \,,
\\
& \bra{\xi_{\nb}} \slashed{\cB}_{n\perp} \chi_n \ket{\epsilon_1^a} 
=
    \parbox{20mm} {\resizebox{20mm}{!}{
  \begin{fmfgraph*}(30,20)
   \fmfset{arrow_len}{3mm}
    \fmfset{arrow_ang}{20}
    \fmfstraight
    \fmfbottom{i1,d1,o1}
    \fmftop{i2,d2,o2}
    \fmflabel{$\epsilon_1^a(k_1)$}{i1}
    \fmflabel{$\xi_{\nb}(p_2)$}{o1}
    \fmf{phantom}{i2,v2}
    \fmf{phantom}{v2,o2}
    \fmf{fermion}{v1,o1}
     \fmf{gluon}{v1,i1}
    \fmffreeze
    \fmf{plain}{v1,i1}
    \fmf{dbl_dots_arrow,f=(0.8,,0.1,,0.1),label=$q_\perp$}{v2,v1}
    \fmfv{d.sh=circle, d.fi=full, d.si=2thick, f=(0.8,,0.1,,0.1)}{v1}
  \end{fmfgraph*}}}
     =  T^a \bar{\xi}_{\nb}\left(\slashed{\epsilon}_{1\perp}-\slashed{k}_{1\perp} \frac{ n\cdot \epsilon_1}{n \cdot k_1}\right) \,.
\end{align}
The tree-level amplitude is then
\begin{align}
   i \cI_\text{SCET}^0 &=
    \bra{\xi_{\nb}; \epsilon_2^b}
    \bar{\chi}_{n}\slashed{\cB}_{n\perp} \frac{1}{\slashed{\cP}_\perp} \cO_{s} \frac{1}{\slashed{\cP}_\perp} \slashed{\cB}_{\nb \perp} \chi_{\nb} 
    \ket{\xi_n; \epsilon_1^a}  
    =
    \begin{gathered}  \parbox{30mm} {\resizebox{30mm}{!}{
    \begin{fmfgraph*}(50,20)
    \fmfstraight
    \fmfleft{i1,i2}
    \fmfright{o1,o2}
    \fmflabel{$\bar{n}$}{i1}
    \fmflabel{$\bar{n}$}{o1}
    \fmflabel{$n$}{o2}
    \fmflabel{$n$}{i2}
    \fmf{gluon}{i1,v1}
    \fmf{gluon}{v2,o2}
    \fmf{fermion}{i2,v2}
    \fmf{fermion}{v1,o1}
    \fmffreeze
    \fmf{dbl_dots_arrow,f=(0.8,,0.1,,0.1),label=$q_\perp$}{v2,v1}
    \fmf{plain}{i1,v1}
    \fmf{plain}{v2,o2}
    \fmfv{d.sh=circle, d.fi=full, d.si=2thick, f=(0.8,,0.1,,0.1)}{v1}
    \fmfv{d.sh=circle, d.fi=full, d.si=2thick, f=(0.8,,0.1,,0.1)}{v2}
    \end{fmfgraph*}}}
\end{gathered}
% \quad\quad
% \equiv\begin{gathered} \parbox{30mm} {\resizebox{30mm}{!}{
%     \begin{fmfgraph*}(50,20)
%     \fmfleft{i1,i2}
%     \fmfright{o1,o2}
%     \fmflabel{$\bar{n}$}{i1}
%     \fmflabel{$\bar{n}$}{o1}
%     \fmflabel{$n$}{o2}
%     \fmflabel{$n$}{i2}
%     \fmf{gluon}{i1,v1}
%     \fmf{gluon}{v1,o2}
%     \fmf{fermion}{i2,v1}
%     \fmf{fermion}{v1,o1}
%     \fmffreeze
%     \fmf{plain}{i1,v1}
%     \fmf{plain}{v1,o2}
%     \fmfv{d.sh=circle, d.fi=full, d.si=3thick, f=(0.8,,0.1,,0.1)}{v1}
%   \end{fmfgraph*}}}
%\end{gathered}
\nonumber \\*[7mm]
&=-i g^2 T^a T^b \xi_{\nb}\left[\slashed{\epsilon}_{1\perp}-\slashed{k}_{1\perp}\frac{n \cdt \epsilon_{1}}{n \cdt k_1}\right]\frac{\slashed{q}_\perp}{t}\left[\slashed{\epsilon}_{2\perp}-\slashed{k}_{2\perp}\frac{\nb \cdt \epsilon_{2}}{\nb \cdt k_2}\right]\xi_{n}\,, \nonumber \\
&= i g^2 T^a T^b \cM_T \,,
\end{align}
in perfect agreement with QCD for all polarizations, so that $C_T =1$ in Eq.~\eqref{OFG} at tree-level.

%% file: diagrams/scattfig.tex
\begin{figure}
\centering
\begin{tikzpicture}
    \node[single arrow, fill=darkerblue,
      minimum width = 10pt, single arrow head extend=3pt,
      minimum height=20mm,
      rotate=17] at (1,2) {}; 
    \node[single arrow, fill=darkred,
      minimum width = 10pt, single arrow head extend=3pt,
      minimum height=20mm,
      rotate=-17] at (4,2) {}; 
    \node[single arrow, fill=darkerblue,
      minimum width = 10pt, single arrow head extend=3pt,
      minimum height=20mm,
      rotate=180-17] at (1,3) {}; 
     \node[single arrow, fill=darkred,
      minimum width = 10pt, single arrow head extend=3pt,
      minimum height=20mm,
      rotate=180+17] at (4,3) {}; 
     \draw [thick, ->] (0,1) -- (2,1) node[midway,above] {$\vec{n}$};
     \draw [thick, ->] (5,1) -- (3,1) node[midway,above] {$-\vec{n}$};
     \draw [thick, ->] (-1,2.3) -- (-1,3.3) node[midway,left] {$q_T$};
     \node[rotate=15] at (0.2,2.3) {$\xi_n(p_1)$};
     \node[rotate=-15] at (4.8,2.3) {$\epsilon_2(k_2)$};
     \node[rotate=-15] at (0.2,3.7) {$\xi_{\nb}(p_2)$};
     \node[rotate=15] at (4.8,3.7) {$\epsilon_1(k_1)$};
\end{tikzpicture}
    \caption{Quark-gluon scattering at small $q_T$. The incoming {\color{darkblue} quark} and outgoing {\color{darkred} gluon} are aligned close to the $n^\mu$ direction and the incoming  {\color{darkred} gluon} and outgoing {\color{darkblue} quark} are alighned close to the $\nb^\mu$ direction. }
    \label{fig:scatfig}
\end{figure}
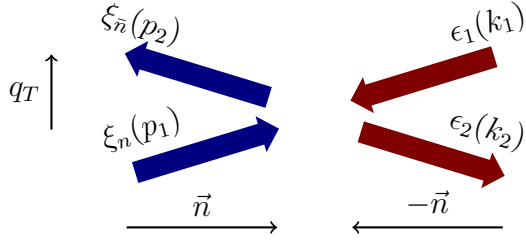

%% file: oneloop.tex
\section{One-loop matching \label{sec:loop}}
Now we proceed to matching at one-loop. We use Feynman gauge throughout, and dimensional regularization in $d=4-2\varepsilon$ dimensions for both UV and IR divergences. Techniques for separating UV and IR divergences, at least at one-loop, are standard. 
In QCD, the separation of UV and IR can be automated -- it is an option in {\tt Package X}~\cite{packageX}, for example. The SCET graphs can have rapidity divergences which we regulate with the $\eta$-regulator, as discussed in Section~\ref{etaregsec}. We always expand in $\eta$ before expanding in $\varepsilon$.
Our analysis extends that of~\cite{MSSV}, where only $\cM_{T}$ was considered.

\subsection{QCD}
\begin{figure}[t]
    \centering
    \includegraphics[scale=0.8]{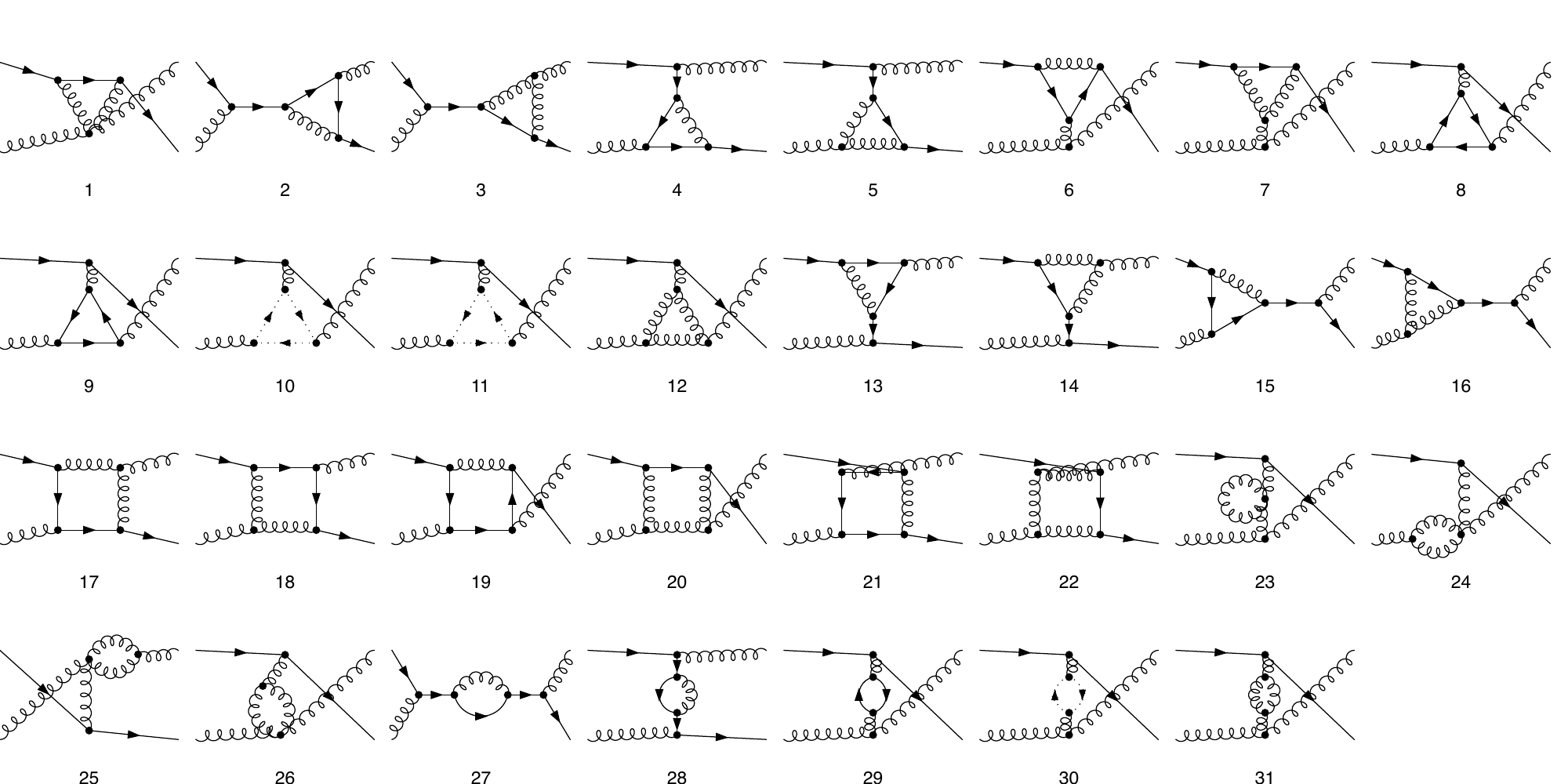}
    \caption{Amputated one-loop diagrams for $q g \to q g$ in QCD. Diagrams 10, 11 and 30 involve ghosts.}
    \label{fig:loopqcd}
\end{figure}

The QCD calculation involves standard methods. In fact, the one-loop matrix elements for quark-gluon scattering are well known (for example, see Refs.~\cite{Ellis:1985er,Fuhrer:2010eu}). For completeless, we reproduced the result from scratch. We generated the diagrams using FeynArts~\cite{Hahn:2000kx}, and evaluated them using Feyncalc~\cite{Shtabovenko:2020gxv} and Package X~\cite{packageX} with the Feynhelpers interface~\cite{Shtabovenko:2016whf}. The diagrams are shown in Figure~\ref{fig:loopqcd}.
We first computed the exact matrix element, then substituted in Eq.~\eqref{ueom} and Eq.~\eqref{Aeom} and expanded to leading power in the Regge limit. 
The result for the one-loop amplitude (excluding external leg corrections) is
\begin{align}
  &\cI_{\text{QCD}}^1=   
    \frac{g^4}{16\pi^2 t}C_F\!\left[\frac{1}{\epsuv}-\frac{2}{\epsir^2}-\frac{4+2\ln{\frac{\mu^2}{Q^2}}}{\epsir}-2\ln{\frac{\mu^2}{Q^2}}\logq+\ln^2{\frac{\mu^2}{-t}}-3\logq+\frac{7\pi^2}{6}-9\right]\!\cM_{T} (T^a T^b)_{ij} \nonumber\\*
    &+\frac{g^4}{16\pi^2 t}C_A\left[\frac{2}{\epsuv}-\frac{2}{\epsir^2}-\frac{2+2\logq}{\epsir}-\ln^2{\frac{\mu^2}{-t}}+\frac{\pi^2}{6}+2\right]\cM_{T}(T^a T^b)_{ij}\ \nonumber\\*
    &-\frac{g^4}{16\pi^2 t}(C_A-C_F)(2\cM_{L1}+2\cM_{L2})(T^a T^b)_{ij} \nonumber\\*
    &+\frac{ g^4}{16\pi^2 t} \delta^{ab}\delta_{ij} \left[\frac{1}{\epsir}+\logq\right](i\pi) \cM_{T} \ .
    \label{IQCD1}
\end{align}
for an $SU(N)$ gauge theory with quarks in the fundamental representation, such as QCD. Here $i$ is the color of the outgoing quark, $j$ the color of the incoming quark, $b$ the color of the outgoing gluon, and $a$ the color of the incoming gluon. Eq.~\eqref{IQCD1} is valid for any color representation of a generic gauge group if one replaces
\begin{align}
\delta^{ab} \delta_{ij} &\to 2 f_{acg}f_{bch}\left\{T^g, T^h\right\}_{ij} -  C_A \left\{T^a,T^b\right\}_{ij}  \,,
\end{align}
(see Ref.~\cite{Fuhrer:2010eu}).
The result for QED is given by the replacements $T^a T^b \to 1$ and $\delta^{ab} \delta_{ij} \to 0$.

The renormalized on-shell $S$-matrix element is given by including the renormalization counterterms, and the renormalized on-shell external-leg corrections needed in the LSZ reduction formula.
The $\overline{\text{MS}}$ wavefunction renormalization counterterms for the quark $Z_{\psi}$ and the gluon $Z_A$ in Feynman gauge are
\begin{align}
    Z_{\psi}&=1-\frac{g^2}{16\pi^2}\frac{C_F}{\epsuv}\ , & Z_{A}&=1+\frac{g^2}{16\pi^2\epsuv}\left[\frac{5C_A}{3}-4n_f T_F\right]\,.
    \label{4.3}
\end{align}
The counterterm $Z_g$ for the coupling constant $g$ is,
\begin{align}
    Z_g=1-\frac{g^2}{32\pi^2\epsuv}\left[\frac{11C_A}{3}-4n_f T_F\right]\,.
    \label{4.4}
\end{align}
The counterterm contribution arises from two insertions of the vertex counterterm $\delta_g+\delta_\psi + \delta_A/2$ at each vertex, and $-\delta_\psi$ from the fermion wavefunction counterterm for the internal fermion propagator, resulting in a net counterterm contribution
\begin{align}
    \cI_{ct}^1 &= \left(2 \delta_g + \delta_\psi + \delta_A  \right) g^2 T^a T^b \cM_T\,,
\end{align}
where $Z_i = 1 + \delta_i $. These counterterms exaclty cancel the UV divergences in $\cI_{\text{QCD}}^1$. 

To use the amplitude for an $S$-matrix element, we must convert to the on-shell scheme. The $\overline{\text{MS}}$-renormalized external-leg correction is
\begin{align}
    \cI_{\mathcal{R}}^1 &= \left(\delta \mathcal{R}_\psi + \delta \mathcal{R}_A  \right) g^2 T^a T^b \cM_T\,,
\end{align}
where $\mathcal{R}_i$ is the renormalized residue of the pole of the particle $i$ propagator, and $\mathcal{R}_i = 1 + \delta \mathcal{R}_i$. Since the on-shell self-energy graphs are scaleless, the $\mathcal{R}_i$ are
\begin{align}
    \mathcal{R}_{\psi}&=1+\frac{g^2}{16\pi^2}\frac{C_F}{\epsir}\ , & \mathcal{R}_{A}&=1-\frac{g^2}{16\pi^2\epsir}\left[\frac{5C_A}{3}-4n_f T_F\right]\,.
    \label{4.5}
\end{align}
The renormalized on-shell $S$-matrix amplitude is then given by 
\begin{align}
    &S_{\text{QCD,ren}}^1= \cI_{\text{QCD}}^1 + \cI_{ct}^1 + \cI_{\mathcal{R}}^1 \nonumber \\
    &=\frac{g^2}{16\pi^2}C_F\left[-\frac{2}{\epsir^2}-\frac{3+2\ln{\frac{\mu^2}{Q^2}}}{\epsir}-2\ln{\frac{\mu^2}{Q^2}}\logq+\ln^2{\frac{\mu^2}{-t}}-3\logq+\frac{7\pi^2}{6}-9\right]\cM_{T} (T^a T^b)_{ij} \nonumber\\
    &+\frac{g^2}{16\pi^2}C_A\left[-\frac{2}{\epsir^2}-\frac{1}{\epsir}\left(\frac{11}{3}+\logq\right)-\ln^2{\frac{\mu^2}{-t}}+\frac{\pi^2}{6}+2\right]\cM_{T}(T^a T^b)_{ij} 
    +\frac{g^2}{16\pi^2}\frac{4 n_f T_F}{\epsir}\cM_{T}(T^a T^b)_{ij} \ \nonumber\\
    &-\frac{g^2}{16\pi^2}(C_A-C_F)(2\cM_{L1}+2\cM_{L2})(T^a T^b)_{ij} \nonumber\\
    &+\frac{ g^4}{16\pi^2} \delta^{ab}\delta_{ij} \left[\frac{1}{\epsir}+\logq\right](i\pi)\cM_{T} \,,
\end{align}
and is UV finite. Since Eq.~\eqref{4.3} and Eq.~\eqref{4.4} are the same in QCD and SCET, we can compute the matching condition between the two theories using the on-shell amplitude Eq.~\eqref{IQCD1} without the renormalization and external-leg factors.

\subsection{SCET}

\input{diagrams/collsoftsumm}

\subsubsection{Collinear graphs}
The first collinear graph is
\begin{align}
    i\cI_{WQ,n} &= 
    \begin{gathered} \parbox{30mm} {\resizebox{30mm}{!}{
 \begin{fmfgraph*}(40,30)
        \fmfleft{i1,i2}
        \fmfright{o1,o2}
        % \fmflabel{$n$}{i2}
        % \fmflabel{$n$}{o2}
        % \fmflabel{$\bar{n}$}{i1}
        % \fmflabel{$\bar{n}$}{o1}
        \fmf{gluon}{i1,v1}
        \fmf{fermion}{v1,o1}
        \fmf{fermion,tension=2.5}{i2,v2}
        \fmf{fermion,label=$p_1+k$}{v2,v1}
        \fmf{gluon,left,tension=0.4,label=$k$}{v2,v1}
        \fmf{gluon}{v1,o2}
        \fmf{fermion,tension=0}{v2,v1}
    \fmffreeze
    \fmf{plain}{i1,v1}
    \fmf{plain,left,tension=0.4}{v2,v1}
    \fmf{plain}{v1,o2}
    \fmfv{d.sh=circle, d.fi=full, d.si=3thick, f=(0.8,,0.1,,0.1)}{v1}
    \end{fmfgraph*}}}
\end{gathered} \nonumber \\
&=    -g^4 T^a T^b \bar{\xi}_{\nb}\left[\slashed{\epsilon}_1-\slashed{k}_{1\perp}\frac{n \cdt \epsilon_{1}}{n \cdt k_1}\right]\frac{1}{\slashed{q}_\perp}
\tilde{\mu}^{2\epsilon}\int \dlipsred{k}{d} \nureg{\nb\cdt k}\frac{1}{[k^2+i0^+][(p_1 +k)^2+i0^+]}
\nonumber \\
& \times\bigg\{\left(C_F-\frac{C_A}{2}\right)\bigg[\frac{1+z}{z}\left(2 \slashed{\epsilon}_{2\perp}+\frac{\slashed{q}_\perp}{Q}\nb \cdt \epsilon_{2}\right)
+\frac{2 +(d-2)z}{{2}z}\left(\frac{z\slashed{q}_\perp-2\slashed{k}_\perp}{Q}\right)\nb\cdot\epsilon_{2} \bigg] \nonumber \\
 & \hspace{3cm} +C_A\bigg[\frac{\slashed{q}_\perp}{Q}+\frac{d-2}{4}\left(\frac{z\slashed{q}_\perp-2\slashed{k}_\perp}{Q}\right)\bigg]\nb \cdt \epsilon_{2}\bigg\}\xi_{n}
\label{IWQ}
\end{align}
where
\begin{equation}
     z = \frac{\nb\cdot k}{Q}\,.
\end{equation}
The graph is scaleless, but being careful to separate UV and IR, it evaluates to
\begin{multline}
    \cI_{WQ,n} =\frac{g^4}{16\pi^2}  T^a T^b \left(C_F-\frac{C_A}{2}\right)\left(\frac{1}{\epsuv}-\frac{1}{\epsir}\right)
    \left[\frac{1}{\eta}+\logp+1\right]\ 2\cM_{T}  \\
     +\frac{g^4}{16\pi^2} T^a T^b C_A\left(\frac{1}{\epsuv}-\frac{1}{\epsir}\right)\left(\cM_{1\nb}+\frac{1}{2}\cM_{q12}\right)\ .
\end{multline}

The gluon Wilson line graph is
\begin{align}
    i\cI_{WG,n} &= 
    \begin{gathered} \parbox{30mm} {\resizebox{30mm}{!}{
 \begin{fmfgraph*}(40,30)
    \fmfleft{i1,i2}
        \fmfright{o1,o2}
        \fmflabel{$n$}{i2}
        \fmflabel{$n$}{o2}
        \fmflabel{$\bar{n}$}{i1}
        \fmflabel{$\bar{n}$}{o1}
        \fmf{gluon}{i1,v1}
        \fmf{fermion}{v1,o1}
        \fmf{fermion}{i2,v1}
        \fmf{gluon,label=$k_2+k$}{v1,v2}
        \fmf{gluon,tension=2.5}{v2,o2}
        \fmf{gluon,left,tension=0.4,label=$k$}{v1,v2}
    \fmffreeze
    \fmf{plain}{i1,v1}
    \fmf{plain}{v1,v2}
    \fmf{plain,tension=2.5}{v2,o2}
    \fmf{plain,left,tension=0.4}{v1,v2}
    \fmfv{d.sh=circle, d.fi=full, d.si=3thick, f=(0.8,,0.1,,0.1)}{v1}
    \end{fmfgraph*}}}
\end{gathered} \nonumber \\
    & =-\frac{1}{2}g^4  T^a T^b \bar{\xi}_{\nb}
    \left[\slashed{\epsilon}_1-\slashed{k}_{1\perp}\frac{n \cdt \epsilon_{1}}{n \cdt k_1}\right]
    \frac{1}{\slashed{q}_\perp} \tilde{\mu}^{2\epsilon}
   \int \dlipsred{k}{d}\frac{1}{k^2(k_2+k)^2}\frac{1}{z(1+z)} \nonumber \\
   & \times
   C_A
   \bigg[(z^2+z+1)\slashed{\epsilon}_{2\perp}
    +\frac{4z^2+3z+2}{4}\frac{\slashed{q}_\perp}{Q}\nb \cdt \epsilon_{2} +\frac{2z+1}{2}\frac{\slashed{k}_\perp}{Q}\nb \cdt \epsilon_{2}\bigg]u_n(p_1)\,.
\end{align}
The factor of $1/2$ out front is a symmetry factor. This graph is also scaleless and evaluates to
\begin{align}
    \cI_{WG,n}&=\frac{g^4}{16\pi^2} T^a T^b C_A\left(\frac{1}{\epsuv}-\frac{1}{\epsir}\right)
    \left[\frac{1}{\eta}+\logp+\frac{1}{2} \right]\cM_{T}\,.
\end{align}
The remaining two collinear graphs are significantly more complicated. The quark V graph is
\vspace{5mm}
\begin{align}
 i\cI_{VQ,n} &=
    \begin{gathered}  {\resizebox{30mm}{!}{
 \begin{fmfgraph*}(40,15)
        \fmfbottom{i1,d1,o1}
        \fmftop{i2,d2,o2}
        \fmflabel{$n$}{i2}
        \fmflabel{$n$}{o2}
        \fmflabel{$\bar{n}$}{i1}
        \fmflabel{$\bar{n}$}{o1}
        \fmf{gluon}{i1,v1}
        \fmf{fermion}{v1,o1}
        \fmf{fermion,tension=2.5}{i2,v2}
        \fmf{fermion,label=$p_1+k$}{v2,v3}
        \fmf{gluon,tension=2.5}{v3,o2}
        \fmf{gluon,tension=0,label=$k$,l.d=10}{v2,v1}
        \fmf{fermion,tension=0,label=$k+q$,l.s=left}{v3,v1}
    \fmffreeze
    \fmf{plain}{v3,o2}
    \fmf{plain}{i1,v1}
    \fmf{plain,tension=0}{v2,v1}
    \fmfv{d.sh=circle, d.fi=full, d.si=3thick, f=(0.8,,0.1,,0.1)}{v1}
    \end{fmfgraph*}}}
\end{gathered} \nonumber \\
& =-  g^4  T^a T^b \bar{\xi}_{\nb}\left[\slashed{\epsilon}_1-\slashed{k}_{1\perp}\frac{n \cdt \epsilon_{1}}{n \cdt k_1}\right]
 \frac{1}{\slashed{q}_\perp} \nonumber \\
& \times \left(C_F-\frac{C_A}{2}\right)\tilde{\mu}^{2\epsilon}\int \dlipsred{k}{d} \frac{\mathcal{N}_{VQ} }
    {[k^2+i0^+][(p_1+k)^2+i0^+][(k+q)^2+i0^+]} \xi_n
\end{align}
where the numerator $\mathcal{N}_{VQ}$ is given by\footnote{One can replace $\slashed{k}_\perp \slashed{q}_\perp \to - \vp{k} \cdot \vp{q}$ in the integrand.}
\begin{align}
    \mathcal{N}_{VQ} &=\left[\frac{2\vp{k}^2+2\vp{k}\cdot \vp{q}}{z}-\frac{d-4}{1+z}\left(\vp{k}+\frac{\vp{q}}{2}\right)^2+\left(\frac{d-6}{4}\ z +\frac{d-8}{4}\right)\vp{q}^2\right]\slashed{\epsilon}_{2\perp} \nonumber \\
    &+\frac{1}{z}\left[-2\left(\vec{k}_{\perp}+\vec{q}_{\perp}\right)^{2} \frac{\slashed{k}_{\perp}}{Q}+\frac{\vec{k}_{\perp}^{2} \slashed{q}_{\perp}+\vec{q}_{\perp}^{2} \slashed{k}_{\perp}}{Q}\right] \nb \cdt \epsilon_{2}+\frac{(4-d)}{(1+z)}\left(\vec{k}_{\perp}+\frac{1}{2} \vec{q}_{\perp}\right)^{2} \frac{\slashed{q}_{\perp}+\slashed{k}_{\perp}}{Q} \nb \cdt \epsilon_{2}\nonumber \\
    &+\frac{(6-d)}{4} \vec{q}_{\perp}^{2} \frac{\slashed{k}_{\perp}}{Q} \nb \cdt \epsilon_{2} +\left[\frac{d-2}{2}\ \left(\vec{k}_{\perp}+\vec{q}_{\perp}\right)^{2}+\frac{2-d}{4}\  \vec{q}_{\perp}^{2}-\vec{k}_{\perp} \cdt \vec{q}_{\perp}\right] \frac{\slashed{q}_\perp}{Q} \nb \cdt \epsilon_{2}\nonumber \\
    &+\left[(d-3)z\ \slashed{q}_\perp -6 \slashed{k}_\perp \right]\ \epsilon_{2\perp}\cdt q_\perp +\left[(12-2d+(d-2)z)\slashed{q}_\perp -2(d-2) \slashed{k}_\perp \right]\epsilon_{2\perp}\cdt k_\perp \nonumber \\
    &+\left[\frac{d-2}{2}\ z+1\right](z\slashed{q}_\perp-2\slashed{k}_\perp)\ Q n\cdt \epsilon_2
\end{align}
Note that this graph produces the small $n\cdot\epsilon_2$ polarization.  Evaluating the graph, and 
eliminating the $n\cdot\epsilon_2$ term using equations of motion in Eq.~\eqref{Aeom}, we find that it evaluates to
\begin{multline}
    \cI_{VQ,n}
    =-\frac{g^4}{16\pi^2} T^a T^b \left(C_F-\frac{C_A}{2}\right) \\*
\times \bigg\{   \left[\frac{1}{\eta}\left(\frac{2}{\epsuv}+2\logq\right)
+\frac{3+2\logp}{\epsuv}
    +2\logq\logp+3\logq-\frac{2\pi^2}{3}+8\right]\cM_{T} 
    - 2 \cM_{L1} \bigg\}
%    -\frac{g^2}{16\pi^2}\left(C_F-\frac{C_A}{2}\right)2 \cM_{L1}.
\end{multline}
Note that this graph is IR finite. 
\\

The final collinear graph is
\begin{align}
   i\tilde{\cI}_{VG,n} &= 
    \begin{gathered} \parbox{30mm} {\resizebox{30mm}{!}{
 \begin{fmfgraph*}(40,16)
        \fmfbottom{i1,d1,o1}
        \fmftop{i2,d2,o2}
        \fmflabel{$n$}{i2}
        \fmflabel{$n$}{o2}
        \fmflabel{$\bar{n}$}{i1}
        \fmflabel{$\bar{n}$}{o1}
        \fmf{gluon}{i1,v1}
        \fmf{fermion}{v1,o1}
        \fmf{fermion,tension=2.5}{i2,v2}
        \fmf{gluon,label=$p_1+k$}{v2,v3}
        \fmf{gluon,tension=2.5}{v3,o2}
        \fmf{fermion,tension=0,label=$k$}{v2,v1}
        \fmf{gluon,tension=0,label=$q+k$,l.s=left}{v3,v1}
    \fmffreeze
    \fmf{plain}{v3,o2}
    \fmf{plain}{i1,v1}
    \fmf{plain,tension=0}{v3,v1}
    \fmf{plain}{v2,v3}
    \fmfv{d.sh=circle, d.fi=full, d.si=3thick, f=(0.8,,0.1,,0.1)}{v1}
    \fmfv{d.sh=circle, d.fi=full, d.si=thick }{v2}
    \fmfv{d.sh=circle, d.fi=full, d.si=thick }{v3}
    \end{fmfgraph*}}}
\end{gathered} \nonumber \\*
& =- g^4  T^a T^b \bar{\xi}_{\nb}\left[\slashed{\epsilon}_1-\slashed{k}_{1\perp}\frac{n \cdt \epsilon_{1}}{n \cdt k_1}\right]\frac{1}{\slashed{q}_\perp} 
C_A\tilde{\mu}^{2\epsilon}\int \dlipsred{k}{d} 
   \frac{\mathcal{N}_{GV}}{[k^2+i0^+][(p_1+k)^2+i0^+][(k+q)^2+i0^+]}\xi_n ,
\end{align}
where
\begin{align}
    \mathcal{N}_{GV}&=\left[\frac{1}{z}\left(\vec{k}_{\perp}^{\,2}+\vec{k}_{\perp} \cdt \vec{q}_{\perp}\right)+\left(- \vec{k}_{\perp} \cdt \vec{q}_{\perp}-\frac{1}{2} \vec{k}_{\perp}^{\,2}+\frac{1}{2} (n \cdt k) Q z-\frac{(3 z+2)}{4} \vec{q}_{\perp}^{\,2}\right)\right] \slashed{\epsilon}_{2\perp} \nonumber \\
    &+\left[\frac{\vec{k}_{\perp}^{\,2}-\vec{q}_{\perp}^{\,2}}{2 z}+\frac{(10+d)}{16} {\vec{q}_{\perp}}^{\, 2}+\frac{(d-3)}{2} (n \cdt k)\ Q\right] \frac{\slashed{k}_{\perp}}{Q} \nb \cdt \epsilon_{2} \nonumber \\
    &+\left[\frac{\vec{k}_{\perp}^{2}+\vec{k}_{\perp} \cdt \vec{q}_{\perp}}{z}-\frac{1}{2} \vec{k}_{\perp} \cdt \vec{q}_{\perp}+\frac{20+2 z-d z}{32} \vec{q}_{\perp}^{2}+\frac{(2+2 z-d z)}{4} (n \cdt k)\ Q\right] \frac{\slashed{q}_\perp}{Q} \nb \cdt \epsilon_{2} \nonumber \\
    &+\left[\frac{(3 d-18)}{4} \slashed{k}_{\perp}+\frac{2 z-3 d z+4}{8} \slashed{q}_\perp\right] q_{\perp}\cdt \epsilon_{2\perp} 
    +\ \left[(d-2) \slashed{k}_{\perp}+\frac{2 z-d z+8}{2} \slashed{q}_{\perp}\right] k_{\perp}\cdt \epsilon_{2\perp} \nonumber \\
    &+\left[\frac{(2 d-4) z+d-6}{4} \slashed{k}_{\perp}+\frac{(4-2d)z^2+(6-d)z-4}{8} \slashed{q}_{\perp}\right] Q\ (n\cdt \epsilon_{2}) .
\end{align}
As with the quark V graph, we use the gluon equation of motion to replace the $n\cdot \epsilon_2$ term. The result is
\begin{align}
    {\cI}_{VG,n} &=-\frac{g^4}{16\pi^2} T^a T^b C_A\bigg[\frac{1}{\eta}\left\{\frac{1}{\epsuv}+\logq\right\}+\frac{\logp}{\epsuv}+\frac{1}{\epsir^2}+\frac{\frac{3}{2}+\logq}{\epsir} +\frac{1}{2}\ln^2{\left(\frac{\mu^2}{-t}\right)}+\logq\logp \nonumber \\ 
    & +\frac{3}{2}\logq-\frac{5\pi^2}{12}+3\bigg] \cM_{T} 
 -\frac{g^4C_A}{16\pi^2} T^a T^b\cM_{L1}  \nonumber \\ 
 &-\frac{g^4 C_A}{16\pi^2}  T^a T^b \left[\frac{3}{4\epsuv}-\frac{1}{\epsir}-\frac{1}{4}\logq-1\right]\left(\cM_{1\nb}+\frac{1}{2}\cM_{q12}\right).
\end{align}
Note that, in contrast to the quark V graph, this graph is not IR finite.

Next, consider the T graphs $\cI_{T1,n}, \cI_{T2,n}$:
\begin{align}
    i\cI_{T1,n} &=\begin{gathered} \parbox{30mm} {\resizebox{30mm}{!}{
 \begin{fmfgraph*}(40,20)
    \fmfstraight
    \fmfbottom{i1,d1,o1}
    \fmftop{i2,d2,o2}
    \fmflabel{$n$}{i2}
    \fmflabel{$n$}{o2}
    \fmflabel{$\bar{n}$}{i1}
    \fmflabel{$\bar{n}$}{o1}
    \fmf{fermion}{i2,v2}
    \fmf{fermion}{v1,o1}
    \fmf{gluon}{o2,v2}
    \fmf{gluon}{v1,i1}
    \fmffreeze
    \fmf{plain}{o2,v2}
    \fmf{plain}{v1,i1}
    \fmf{gluon,right=0.4,tension=0.2,label=$k$,l.s=right,l.d=0.085w}{v1,v2}
    \fmf{fermion,right=0.4,tension=0.2,label=$q-k$}{v2,v1}
    \fmf{plain,right=0.4,tension=0.2}{v1,v2}
    \fmfv{d.sh=circle, d.fi=full, d.si=thick}{v2}
    \fmfv{d.sh=circle, d.fi=full, d.si=3thick, f=(0.8,,0.1,,0.1)}{v1}
    \end{fmfgraph*}}}%=0
\end{gathered} \nonumber \\*
& =g^4  T^a T^b\bar{\xi}_{\nb}\left[\slashed{\epsilon}_1-\slashed{k}_{1\perp}\frac{n \cdt \epsilon_{1}}{n \cdt k_1}\right]\frac{1}{\slashed{q}_\perp} \tilde{\mu}^{2\epsilon}\int \dlipsred{k}{d} \frac{1}{[k^2+i0^+][(q-k)^2+i0^+]} \nonumber \\*
& \times \bigg\{\left(C_F-\frac{C_A}{2}\right)(d-4)\left(\frac{\nb\cdt k}{\nb\cdt(p_1-k)}\slashed{\epsilon}_{2\perp}+ \frac{\slashed{q}_\perp-\slashed{k}_\perp}{\nb\cdt (p_1-k)}\nb\cdt\epsilon_2\right) \nonumber \\*
 & \hspace{1cm} +C_F(2-d)\left(\frac{\nb\cdt k}{\nb\cdt q}\slashed{\epsilon}_{2\perp} -\frac{\nb\cdt k}{\nb\cdt p_1}\frac{\slashed{q}_\perp}{\nb\cdt q}\frac{\nb\cdt \epsilon_2}{2}\right)\bigg\}\xi_{n}
\end{align}
and
\begin{align}
    i\cI_{T2,n} &=\begin{gathered} \parbox{30mm} {\resizebox{30mm}{!}{
 \begin{fmfgraph*}(40,25)
    \fmfstraight
    \fmfbottom{i1,d1,o1}
    \fmftop{i2,d2,o2}
    \fmflabel{$n$}{i2}
    \fmflabel{$n$}{o2}
    \fmflabel{$\bar{n}$}{i1}
    \fmflabel{$\bar{n}$}{o1}
    \fmf{fermion,label=$p_1$,l.s=left}{i2,v2}
    \fmf{fermion}{v1,o1}
    \fmf{gluon,label=$k_2$,l.d=10}{o2,v2}
    \fmf{gluon}{v1,i1}
    \fmffreeze
    \fmf{plain}{o2,v2}
    \fmf{plain}{v1,i1}
    \fmf{fermion,label=$p_1-k_2$}{v2,v3}
    \fmf{gluon,right=0.5,tension=0.2,l.s=right,l.d=0.085w}{v1,v3}
    \fmf{fermion,right=0.5,tension=0.2}{v3,v1}
    \fmf{plain,right=0.4,tension=0.2}{v1,v3}
    \fmfv{d.sh=circle, d.fi=full, d.si=thick}{v2}
    \fmfv{d.sh=circle, d.fi=full, d.si=3thick, f=(0.8,,0.1,,0.1)}{v1}
    \end{fmfgraph*}}}
\end{gathered} \nonumber \\
& =g^4 T^a T^b \bar{\xi}_{\nb}\left[\slashed{\epsilon}_1-\slashed{k}_{1\perp}\frac{n \cdt \epsilon_{1}}{n \cdt k_1}\right]\frac{1}{\slashed{q}_\perp} \tilde{\mu}^{2\epsilon}\int \dlipsred{k}{d} \frac{C_F}{[k^2+i0^+][(q-k)^2+i0^+]q_\perp^2} \nonumber \\
& \bigg\{\left(\frac{4-d}{2}\vp{q}^2\frac{\slashed{q}_\perp}{\nb\cdt p_1}+\frac{(2-d)\vp{q}^2 (\nb\cdt k) \slashed{q}_\perp}{2(\nb\cdt p_1)(\nb\cdt q)}\right)\nb\cdt \epsilon_2 \nonumber \\ 
& +(2-d) (\nb\cdt k)\ \slashed{q}_\perp (n\cdt\epsilon_2) +\vp{q}^2\left((d-2)\frac{\nb\cdt k}{\nb\cdt q}+\left(1-\frac{d}{2}\right)\frac{\nb\cdt k}{\nb\cdt q}+d-4\right)\slashed{\epsilon}_{2\perp} \nonumber \\
& -(2-d)\slashed{\epsilon}_{2\perp}\slashed{k}_\perp\slashed{q}_\perp
 +(d-2)\left(2\slashed{k}_\perp -\frac{\nb\cdt k}{\nb\cdt p_1}\slashed{q}_\perp\right)\epsilon_{2\perp}\cdt q_\perp -2\left(d-2\right)\slashed{q}_\perp (k_\perp \cdt \epsilon_{2\perp})\bigg\}
\end{align}
The $k^+ = n\cdot k$ integral in both of these graphs is
\begin{equation}
    \int d k^+ \frac{1}{[k^+ k^- -\vec{k}_\perp^2 + i0^+][k^+k^- -q^+ k^- -(\vec{q}_\perp - \vec{k}_\perp)^2 + i0^+]}\,.
\end{equation}
Since the  $k^+$ poles are on the same side of the real axis these graphs vanish upon contour integration,
\begin{equation}
    \cI_{T1,n} = \cI_{T2,n} = 0 \,.
\end{equation}
More generally, T graphs of this form, where only Glauber momentum $q$ and no collinear momentum flows through a collinear loop, have no large momenta scale for the loop denominators to depend on. Since $q^- = \nb \cdot q =0$, the  only large momentum component that can ever multiply $k^+$ in any of the denominators is $k^-$, and so the poles in the $k^+$ integral will always be on the same side of the real axis and these graphs will always vanish.

The sum of the  $n$-collinear graphs is 
\begin{align}
    &\cI_{n,\text{coll}}=\cI_{VQ,n}+\cI_{VG,n}+\cI_{WG,n}+\cI_{WQ,n}+\cI_{T1,n}+\cI_{T2,n} \nonumber\\
    &=-\frac{g^4}{16\pi^2} T^a T^b C_F\left[\frac{2}{\eta}\left\{\frac{1}{\epsir}+\logq\right\}+\frac{1}{\epsuv}+\frac{2+2\logp}{\epsir}+2\logq\logp+3\logq-\frac{2\pi^2}{3}+8\right]\cM_{T} \nonumber\\
    &-\frac{g^4}{16\pi^2} T^a T^b C_A\left[-\frac{1}{\epsuv}+\frac{1}{\epsir^2}+\frac{1+\logq}{\epsir}+\frac{1}{2}\ln^2{\frac{\mu^2}{-t}}-\frac{\pi^2}{12}-1\right]\cM_{T}\ -\frac{g^4}{16\pi^2}  T^a T^b 2(C_A-C_F)\cM_{L1} \nonumber\\
    &+\frac{g^4}{16\pi^2} T^a T^b C_A \left[\frac{1}{4\epsuv}+\frac{1}{4}\logq+1\right]\left(\cM_{1\nb}+\frac{1}{2}\cM_{q12}\right) \,.
    \label{Incoll}
\end{align}
Excluding the terms on the last line, the collinear contribution by itself satisfies the Ward identity. The last line is troublesome, and does not satisfy the Ward identity. However, as we will see in Section~\ref{sec:zerobin}, it is precisely canceled by zero bin subtractions. 

The $\nb$-collinear graphs give the same results with the replacements $\cM_{L1} \to \cM_{L2}$ and $\cM_{1\nb} \to \cM_{2n}$.

%% file: diagrams/collsoftsumm.tex
For the SCET calculation we need to compute 6 collinear graphs. Four of them were discussed in~\cite{MSSV}
\begin{equation}
 i\cI_{WQ,n}= 
 \begin{gathered} \parbox{20mm} {\resizebox{20mm}{!}{
 \begin{fmfgraph*}(40,20)
        \fmfleft{i1,i2}
        \fmfright{o1,o2}
        \fmflabel{$n$}{i2}
        \fmflabel{$n$}{o2}
        \fmflabel{$\bar{n}$}{i1}
        \fmflabel{$\bar{n}$}{o1}
        \fmf{gluon}{i1,v1}
        \fmf{fermion}{v1,o1}
        \fmf{fermion,tension=2.5}{i2,v2}
        \fmf{fermion}{v2,v1}
        \fmf{gluon,left,tension=0.4}{v2,v1}
        \fmf{gluon}{v1,o2}
        \fmf{fermion,tension=0}{v2,v1}
    \fmffreeze
    \fmf{plain}{i1,v1}
    \fmf{plain,left,tension=0.4}{v2,v1}
    \fmf{plain}{v1,o2}
    \fmfv{d.sh=circle, d.fi=full, d.si=3thick, f=(0.8,,0.1,,0.1)}{v1}
    \end{fmfgraph*}}}
\end{gathered},
\quad
 i\cI_{WG,n}= 
 \begin{gathered} \parbox{20mm} {\resizebox{20mm}{!}{
 \begin{fmfgraph*}(40,20)
        \fmfleft{i1,i2}
        \fmfright{o1,o2}
        \fmflabel{$n$}{i2}
        \fmflabel{$n$}{o2}
        \fmflabel{$\bar{n}$}{i1}
        \fmflabel{$\bar{n}$}{o1}
        \fmf{gluon}{i1,v1}
        \fmf{fermion}{v1,o1}
        \fmf{fermion}{i2,v1}
        \fmf{gluon}{v1,v2}
        \fmf{gluon,tension=2.5}{v2,o2}
        \fmf{gluon,left,tension=0.4}{v1,v2}
    \fmffreeze
    \fmf{plain}{i1,v1}
    \fmf{plain}{v1,v2}
    \fmf{plain,tension=2.5}{v2,o2}
    \fmf{plain,left,tension=0.4}{v1,v2}
    \fmfv{d.sh=circle, d.fi=full, d.si=3thick, f=(0.8,,0.1,,0.1)}{v1}
    \end{fmfgraph*}}}, 
    \end{gathered}
    \quad
    i\cI_{VQ,n} =
    \begin{gathered}  \parbox{20mm} {\resizebox{20mm}{!}{
    \begin{fmfgraph*}(40,16)
        \fmfbottom{i1,d1,o1}
        \fmftop{i2,d2,o2}
        \fmflabel{$n$}{i2}
        \fmflabel{$n$}{o2}
        \fmflabel{$\bar{n}$}{i1}
        \fmflabel{$\bar{n}$}{o1}
        \fmf{gluon}{i1,v1}
        \fmf{fermion}{v1,o1}
        \fmf{fermion,tension=2.5}{i2,v2}
        \fmf{fermion}{v2,v3}
        \fmf{gluon,tension=2.5}{v3,o2}
        \fmf{gluon,tension=0}{v2,v1}
        \fmf{fermion,tension=0}{v3,v1}
    \fmffreeze
    \fmf{plain}{v3,o2}
    \fmf{plain}{i1,v1}
    \fmf{plain,tension=0}{v2,v1}
    \fmfv{d.sh=circle, d.fi=full, d.si=3thick, f=(0.8,,0.1,,0.1)}{v1}
    \end{fmfgraph*}}}
\end{gathered},
\quad
 i\cI_{VG,n}=
\begin{gathered} \parbox{20mm} {\resizebox{20mm}{!}{
 \begin{fmfgraph*}(40,16)
        \fmfbottom{i1,d1,o1}
        \fmftop{i2,d2,o2}
        \fmflabel{$n$}{i2}
        \fmflabel{$n$}{o2}
        \fmflabel{$\bar{n}$}{i1}
        \fmflabel{$\bar{n}$}{o1}
        \fmf{gluon}{i1,v1}
        \fmf{fermion}{v1,o1}
        \fmf{fermion,tension=2.5}{i2,v2}
        \fmf{gluon}{v2,v3}
        \fmf{gluon,tension=2.5}{v3,o2}
        \fmf{fermion,tension=0}{v2,v1}
        \fmf{gluon,tension=0}{v3,v1}
    \fmffreeze
    \fmf{plain}{v3,o2}
    \fmf{plain}{i1,v1}
    \fmf{plain,tension=0}{v3,v1}
    \fmf{plain}{v2,v3}
    \fmfv{d.sh=circle, d.fi=full, d.si=3thick, f=(0.8,,0.1,,0.1)}{v1}
    \fmfv{d.sh=circle, d.fi=full, d.si=thick }{v2}
    \fmfv{d.sh=circle, d.fi=full, d.si=thick }{v3}
    \end{fmfgraph*}}}
\end{gathered}
\end{equation}
In~\cite{MSSV} only the only the rapidity divergent parts of these graphs were computed, and only for the $\epsilon_\perp$ polarizations. 
For the matching we need the entire amplitude, including UV and IR divergent pieces and finite terms. 
There are two additional collinear graphs (``T graphs''),
\begin{align}
    i\cI_{T1,n} &=\begin{gathered} \parbox{20mm} {\resizebox{20mm}{!}{
 \begin{fmfgraph*}(40,20)
    \fmfstraight
    \fmfbottom{i1,d1,o1}
    \fmftop{i2,d2,o2}
    \fmflabel{$n$}{i2}
    \fmflabel{$n$}{o2}
    \fmflabel{$\bar{n}$}{i1}
    \fmflabel{$\bar{n}$}{o1}
    \fmf{fermion}{i2,v2}
    \fmf{fermion}{v1,o1}
    \fmf{gluon}{o2,v2}
    \fmf{gluon}{v1,i1}
    \fmffreeze
    \fmf{plain}{o2,v2}
    \fmf{plain}{v1,i1}
    \fmf{gluon,right=0.4,tension=0.2,label=$n$,l.s=right,l.d=0.085w}{v1,v2}
    \fmf{fermion,right=0.4,tension=0.2,label=$n$}{v2,v1}
    \fmf{plain,right=0.4,tension=0.2}{v1,v2}
    \fmfv{d.sh=circle, d.fi=full, d.si=thick}{v2}
    \fmfv{d.sh=circle, d.fi=full, d.si=3thick, f=(0.8,,0.1,,0.1)}{v1}
    \end{fmfgraph*}}}
\end{gathered}\quad,
&
i\cI_{T2,n}&=\begin{gathered} \parbox{20mm} {\resizebox{20mm}{!}{
 \begin{fmfgraph*}(40,20)
    \fmfstraight
    \fmfbottom{i1,d1,o1}
    \fmftop{i2,d2,o2}
    \fmflabel{$n$}{i2}
    \fmflabel{$n$}{o2}
    \fmflabel{$\bar{n}$}{i1}
    \fmflabel{$\bar{n}$}{o1}
    \fmf{fermion}{i2,v2}
    \fmf{fermion}{v1,o1}
    \fmf{gluon}{o2,v2}
    \fmf{gluon}{v1,i1}
    \fmffreeze
    \fmf{plain}{o2,v2}
    \fmf{plain}{v1,i1}
    \fmf{fermion,label=$n$}{v2,v3}
    \fmf{gluon,right=0.4,tension=0.2,label=$n$,l.s=right,l.d=0.085w}{v1,v3}
    \fmf{fermion,right=0.4,tension=0.2,label=$n$}{v3,v1}
    \fmf{plain,right=0.4,tension=0.2}{v1,v3}
    \fmfv{d.sh=circle, d.fi=full, d.si=thick}{v2}
    \fmfv{d.sh=circle, d.fi=full, d.si=3thick, f=(0.8,,0.1,,0.1)}{v1}
    \end{fmfgraph*}}}\quad .
\end{gathered}
\end{align}
Just the $n$-collinear graphs are shown. The $\nb$-collinear graphs are identical with $n \leftrightarrow \nb$ and must also be included.
We also need two soft graphs
\begin{align}
i \cI_{\text{flower}} &= 
 \begin{gathered} \parbox{20mm} {\resizebox{20mm}{!}{
  \begin{fmfgraph*}(40,25)
    \fmfstraight
    \fmfbottom{i1,d1,o1}
    \fmftop{i2,d3,o2}
    \fmflabel{$n$}{i2}
    \fmflabel{$n$}{o2}
    \fmflabel{$\bar{n}$}{i1}
    \fmflabel{$\bar{n}$}{o1}
    \fmf{fermion}{i2,v3}
    \fmf{fermion}{v1,o1}
    \fmf{gluon}{o2,v3}
    \fmf{gluon}{v1,i1}
    \fmffreeze
    \fmf{plain}{o2,v3}
    \fmf{plain}{v1,i1}
    \fmf{dbl_dots,f=(0.8,,0.1,,0.1)}{v3,v2}
    \fmf{dbl_dots,f=(0.8,,0.1,,0.1)}{v2,v1}
    \fmf{gluon,tension=0.8,foreground=(0,,0.5,,0.0)}{v2,v2}
    \fmfv{d.sh=circle, d.fi=full, d.si=2thick, f=(0.8,,0.1,,0.1)}{v1}
    \fmfv{d.sh=circle, d.fi=full, d.si=2thick, f=(0.8,,0.1,,0.1)}{v2}
    \fmfv{d.sh=circle, d.fi=full, d.si=2thick, f=(0.8,,0.1,,0.1)}{v3}
  \end{fmfgraph*}}}
\end{gathered}\quad ,
&
i\cI_{\text{eye}} &= \begin{gathered} \parbox{20mm} {\resizebox{20mm}{!}{
 \begin{fmfgraph*}(40,25)
    \fmfstraight
    \fmfbottom{i1,d1,o1}
    \fmftop{i2,d4,o2}
    \fmflabel{$n$}{i2}
    \fmflabel{$n$}{o2}
    \fmflabel{$\bar{n}$}{i1}
    \fmflabel{$\bar{n}$}{o1}
    \fmf{fermion}{i2,v4}
    \fmf{fermion}{v1,o1}
    \fmf{gluon}{o2,v4}
    \fmf{gluon}{v1,i1}
    \fmffreeze
    \fmf{plain}{o2,v4}
    \fmf{plain}{v1,i1}
    \fmf{dbl_dots,f=(0.8,,0.1,,0.1)}{v2,v1}
    \fmf{dbl_dots,f=(0.8,,0.1,,0.1)}{v4,v3}
    \fmf{fermion,right,tension=0.2,f=(0.0,,0.5,,0.0)}{v3,v2}
    \fmf{gluon,left,tension=0.2,f=(0.0,,0.5,,0.0)}{v3,v2}
    \fmfv{d.sh=circle, d.fi=full, d.si=2thick, f=(0.8,,0.1,,0.1)}{v1}
    \fmfv{d.sh=circle, d.fi=full, d.si=2thick, f=(0.8,,0.1,,0.1)}{v4}
    \fmfv{d.sh=circle, d.fi=full, d.si=2thick, f=(0.8,,0.1,,0.1)}{v2}
    \fmfv{d.sh=circle, d.fi=full, d.si=2thick, f=(0.8,,0.1,,0.1)}{v3}
  \end{fmfgraph*}}}
 \end{gathered} \quad .
\end{align}
In addition, we have to include appropriate zero bin subtractions. As we will see, the zero bin subtractions are rapidity-finite and nonvanishing in this case, and critical to getting agreement with QCD.

%% file: soft.tex
\subsubsection{Soft graphs}
The soft flower graph arises from Wilson line emissions from both collinear directions. It takes the following form 
\begin{align}
\label{Iflower}
   i \cI_{\text{flower}} &= 
 \begin{gathered}{\resizebox{30mm}{!}{
  \begin{fmfgraph*}(40,25)
    \fmfstraight
    \fmfbottom{i1,d1,o1}
    \fmftop{i2,d3,o2}
    \fmflabel{$n$}{i2}
    \fmflabel{$n$}{o2}
    \fmflabel{$\bar{n}$}{i1}
    \fmflabel{$\bar{n}$}{o1}
    \fmf{fermion}{i2,v3}
    \fmf{fermion}{v1,o1}
    \fmf{gluon}{o2,v3}
    \fmf{gluon}{v1,i1}
    \fmffreeze
    \fmf{plain}{o2,v3}
    \fmf{plain}{v1,i1}
    \fmf{dbl_dots,f=(0.8,,0.1,,0.1)}{v3,v2}
    \fmf{dbl_dots,f=(0.8,,0.1,,0.1)}{v2,v1}
    \fmf{gluon,tension=0.8,foreground=(0,,0.5,,0.0),label=$k$,l.d=15}{v2,v2}
    \fmfv{d.sh=circle, d.fi=full, d.si=2thick, f=(0.8,,0.1,,0.1)}{v1}
    \fmfv{d.sh=circle, d.fi=full, d.si=2thick, f=(0.8,,0.1,,0.1)}{v2}
    \fmfv{d.sh=circle, d.fi=full, d.si=2thick, f=(0.8,,0.1,,0.1)}{v3}
  \end{fmfgraph*}}}
\end{gathered}  \nonumber \\
&    = -2 g^4  T^a T^b C_F \bar{\xi}_{\nb}\left[\slashed{\epsilon}_1-\slashed{k}_{1\perp}\frac{n \cdt \epsilon_{1}}{n \cdt k_1}\right]\frac{1}{t} \nonumber \\*
&    \times\tilde{\mu}^{2\epsilon}\left[\int\dlipsred{k}{d} \nureg{|2k_z|}\frac{\slashed{q}_\perp+\slashed{k}_\perp}{(\nb\cdot k)(n\cdot k)\ (k^2+i0^+) }\right]\left[\slashed{\epsilon}_2-\slashed{k}_{2\perp}\frac{\nb \cdt \epsilon_{2}}{\nb \cdt k_2}\right]\xi_{n} \,.
\end{align}
The graph is scaleless and evaluates to
\begin{equation}
    \cI_\text{flower} =\frac{ g^4}{16\pi^2} T^a T^b C_F
    \left[ \left(-\frac{4}{\eta} + 2 \ln{\frac{\mu^2}{\nu^2}} \right)\left(\frac{1}{\epsuv}-\frac{1}{\epsir}\right)
    +\frac{2}{\epsuv^2}-\frac{2}{\epsir^2}\right]\cM_{T}\,.
\end{equation}

The eye graph $\cI_{\text{eye}}$ arises from a time ordered product of the soft-Glauber operators $\cO_{S n}$ and  $\cO_{S \nb}$  in \eqref{eq:softglauber},
\begin{align}\label{eq:eye}
  & i\cI_{\text{eye}} =\hspace{1cm} \begin{gathered} {\resizebox{15mm}{!}{
 \begin{fmfgraph*}(20,25)
    \fmfstraight
    \fmfbottom{i1,d1,o1}
    \fmftop{i2,d4,o2}
    \fmflabel{$n$}{i2}
    \fmflabel{$n$}{o2}
    \fmflabel{$\bar{n}$}{i1}
    \fmflabel{$\bar{n}$}{o1}
    \fmf{fermion}{i2,v4}
    \fmf{fermion}{v1,o1}
    \fmf{gluon}{o2,v4}
    \fmf{gluon}{v1,i1}
    \fmffreeze
    \fmf{plain}{o2,v4}
    \fmf{plain}{v1,i1}
    \fmf{dbl_dots,f=(0.8,,0.1,,0.1)}{v2,v1}
    \fmf{dbl_dots,f=(0.8,,0.1,,0.1)}{v4,v3}
    \fmf{fermion,right,tension=0.2,f=(0.0,,0.5,,0.0),label=$k+q$}{v3,v2}
    \fmf{gluon,left,tension=0.2,f=(0.0,,0.5,,0.0),label=$k$}{v3,v2}
    \fmfv{d.sh=circle, d.fi=full, d.si=2thick, f=(0.8,,0.1,,0.1)}{v1}
    \fmfv{d.sh=circle, d.fi=full, d.si=2thick, f=(0.8,,0.1,,0.1)}{v4}
    \fmfv{d.sh=circle, d.fi=full, d.si=2thick, f=(0.8,,0.1,,0.1)}{v2}
    \fmfv{d.sh=circle, d.fi=full, d.si=2thick, f=(0.8,,0.1,,0.1)}{v3}
  \end{fmfgraph*}}}
 \end{gathered} \nonumber \\
     & = - g^4 T^a T^b C_F 
 \bar{\xi}_{\nb}\left[\slashed{\epsilon}_1-\slashed{k}_{1\perp}\frac{n \cdt \epsilon_{1}}{n \cdt k_1}\right]\frac{1}{\slashed{q}_\perp}
 \nonumber\\
    &\times \tilde{\mu}^{2\epsilon} \left[\int\dlipsred{k}{d} \nureg{|2k_z|}\frac{1}{(k^2+i0^+)\left[(k+q)^2+i0^+\right]}\left(\gamma^\mu_\perp (\slashed{k} +\slashed{q}_\perp)\gamma_{\mu\perp} + \frac{2\ \slashed{k}_\perp(\slashed{k} +\slashed{q}_\perp)\slashed{k}_\perp}{(\nb\cdot k)(n\cdot k)}\right)\right]\nonumber\\
    & \hspace{2cm} \times \frac{1}{\slashed{q}_\perp}\left[\slashed{\epsilon}_2-\slashed{k}_{2\perp}\frac{\nb \cdt \epsilon_{2}}{\nb \cdt k_2}\right]\xi_{n}\ .
\end{align}
Evaluating the graph, we find that
\begin{align}
    \cI_{\text{eye}}
    &= \frac{2 g^4}{16\pi^2} T^a T^b C_F \bigg[\frac{1}{\eta}
    \left(\frac{2}{\epsuv}+2\logq\right)-\frac{1}{\epsuv^2}
    -\frac{\ln{\frac{\mu^2}{\nu^2}}}{\epsuv}  
    +\frac{3}{2\epsuv}-\frac{1}{2}\ln^2{\frac{\mu^2}{-t}}+\logq\ln{\frac{\nu^2}{-t}} \nonumber \\ 
    & \hspace{1cm} +\frac{3}{2}\logq-\frac{\pi^2}{12}+\frac{7}{2} \bigg]\cM_{T} \,.
    \label{eyegraph}
\end{align}
The full analytic form for the soft eye graph is given in Appendix~\ref{sec:eye}.
The sum of the two soft graphs is
\begin{align}
    \cI_\text{soft} &=\cI_{\text{eye}}+\cI_{\text{flower}} \nonumber \\
&    =\frac{2g^2C_F}{16\pi^2} T^a T^b \bigg[\frac{1}{\eta}\left(\frac{2}{\epsir}+2\logq\right)-\frac{1}{\epsir^2}-\frac{\ln{\frac{\mu^2}{\nu^2}}}{\epsir}+\frac{3}{2\epsuv}
    -\frac{1}{2}\ln^2{\frac{\mu^2}{-t}}+\logq\ln{\frac{\nu^2}{-t}} \nonumber \\
    & \hspace{1cm} +\frac{3}{2}\logq-\frac{\pi^2}{12}+\frac{7}{2} \bigg]\cM_{T}\,.
    \label{Isoft}
\end{align}
As a non-trivial check on our results, we find that the soft sector of SCET$_{\text{II}}$ is gauge invariant on its own, as should be the case. 

%% file: boxes.tex
\subsubsection{Glauber Boxes}
Next, we consider Glauber box diagrams. Since Glauber fermion exchange flips quark $\leftrightarrow$ gluon, there must be an odd number of fermionic Glauber exchanges in the Regge limit (backwards scattering) for $qg \to qg$. So at one-loop the box graphs must have one fermionic Glauber and one gluonic Glauber. The allowed graphs are
\input{diagrams/glauberboxes}
\vspace{0.5cm}

\noindent Although these graphs are not rapidity divergent, one still must use a a rapidity regulator like the $\eta$ regulator to evaluate the graphs graphs. The $\eta$ regulator inserts factors of $\nureg{|2k_z|}$ in the loop integrand which makes the integrals well defined.  

The last two (crossed-box) graphs vanish using the $\eta$ regulator.  The first two graphs have been worked out in~\cite{MSSV} and contribute pure imaginary pieces, and we simply restate the result here:
\begin{equation}
\label{Iboxes}
    \cI_{\text{boxes}}=\frac{ g^4}{16\pi^2} \delta^{ab}\delta_{ij} \left[\frac{1}{\epsir}+\logq\right](i\pi)\ .
\end{equation}
That the Glauber box graph give a purely imaginary contribution is consistent with general observations about Glauber gluons: the $i\pi$ is present even in QED for processes like $\gamma \gamma \to e^+e^-$, where it represents the leading order expansion of the Coulomb phase. For a crossed process where only one incoming and one outgoing particle is charged, like $\gamma e^- \to \gamma e^-$ the phase is absent. Because the $i\pi$ disappears upon crossing, in order for graphs like the Glauber boxes to give the pure phase, they must have a non-analytic integrand, as imposed by the $|2k_z|$ factor in the $\eta$-regulator (this is not true in full QCD where the $i\pi$ comes with an associated logarithm). A further discussion of the essential non-analyticity of rapidity regularization can be found in~\cite{mattcollinearviolation}.

%% file: diagrams/glauberboxes.tex
%Glauber Boxes Figures
\begin{align}
&\begin{gathered} \parbox{30mm} {\resizebox{30mm}{!}{
\begin{fmfgraph*}(40,18)
 \fmfset{arrow_len}{3mm}
 \fmfset{arrow_ang}{20}
\fmfstraight
\fmfbottom{i1,d1,o1}
\fmftop{i2,d2,o2}
\fmflabel{$n$}{i2}
\fmflabel{$n$}{o2}
\fmflabel{$\bar{n}$}{i1}
\fmflabel{$\bar{n}$}{o1}
\fmf{fermion}{i2,v2}
\fmf{fermion}{v2,v3}
\fmf{gluon}{v3,o2}
\fmf{gluon}{i1,v1}
\fmf{gluon}{v1,v4}
\fmf{fermion}{v4,o1}
\fmffreeze
\fmf{plain}{v3,o2}
\fmf{plain}{i1,v1}
\fmf{plain}{v1,v4}
\fmf{dbl_dots,f=(0.8,,0.1,,0.1),tension=0}{v2,v1}
\fmf{dbl_dots_arrow,f=(0.8,,0.1,,0.1),tension=0}{v3,v4}
\fmfv{d.sh=circle, d.fi=full, d.si=3thick, f=(0.8,,0.1,,0.1)}{v2}
\fmfv{d.sh=circle, d.fi=full, d.si=3thick, f=(0.8,,0.1,,0.1)}{v1}
\fmfv{d.sh=circle, d.fi=full, d.si=3thick, f=(0.8,,0.1,,0.1)}{v3}
\fmfv{d.sh=circle, d.fi=full, d.si=3thick, f=(0.8,,0.1,,0.1)}{v4}
\end{fmfgraph*}}}
\end{gathered} &
&\begin{gathered}  \parbox{30mm} {\resizebox{30mm}{!}{
 \begin{fmfgraph*}(40,18)
 \fmfset{arrow_len}{3mm}
 \fmfset{arrow_ang}{20}
\fmfstraight
\fmfbottom{i1,d1,o1}
\fmftop{i2,d2,o2}
\fmflabel{$n$}{i2}
\fmflabel{$n$}{o2}
\fmflabel{$\bar{n}$}{i1}
\fmflabel{$\bar{n}$}{o1}
\fmf{fermion}{i2,v2}
\fmf{gluon}{v2,v3}
\fmf{gluon}{v3,o2}
\fmf{gluon}{i1,v1}
\fmf{fermion}{v1,v4}
\fmf{fermion}{v4,o1}
\fmffreeze
\fmf{plain}{v2,v3}
\fmf{plain}{v3,o2}
\fmf{plain}{i1,v1}
\fmf{dbl_dots_arrow,f=(0.8,,0.1,,0.1),tension=0}{v2,v1}
\fmf{dbl_dots,f=(0.8,,0.1,,0.1),tension=0}{v3,v4}
\fmfv{d.sh=circle, d.fi=full, d.si=3thick, f=(0.8,,0.1,,0.1)}{v2}
\fmfv{d.sh=circle, d.fi=full, d.si=3thick, f=(0.8,,0.1,,0.1)}{v1}
\fmfv{d.sh=circle, d.fi=full, d.si=3thick, f=(0.8,,0.1,,0.1)}{v3}
\fmfv{d.sh=circle, d.fi=full, d.si=3thick, f=(0.8,,0.1,,0.1)}{v4}
\end{fmfgraph*}}}
\end{gathered} &
&\begin{gathered} \parbox{30mm} {\resizebox{30mm}{!}{
\begin{fmfgraph*}(40,18)
 \fmfset{arrow_len}{3mm}
 \fmfset{arrow_ang}{20}
\fmfstraight
\fmfbottom{i1,d1,o1}
\fmftop{i2,d2,o2}
\fmflabel{$n$}{i2}
\fmflabel{$n$}{o2}
\fmflabel{$\bar{n}$}{i1}
\fmflabel{$\bar{n}$}{o1}
\fmf{fermion}{i2,v2}
\fmf{fermion}{v2,v3}
\fmf{gluon}{v3,o2}
\fmf{gluon}{i1,v1}
\fmf{fermion}{v1,v4}
\fmf{fermion}{v4,o1}
\fmffreeze
\fmf{plain}{v3,o2}
\fmf{plain}{i1,v1}
\fmf{phantom,tension=0}{v3,v1}
\fmf{dbl_dots,f=(0.8,,0.1,,0.1),tension=0}{v2,v4}
\fmfv{d.sh=circle, d.fi=full, d.si=3thick, f=(0.8,,0.1,,0.1)}{v2}
\fmfv{d.sh=circle, d.fi=full, d.si=3thick, f=(0.8,,0.1,,0.1)}{v1}
\fmfv{d.sh=circle, d.fi=full, d.si=3thick, f=(0.8,,0.1,,0.1)}{v3}
\fmfv{d.sh=circle, d.fi=full, d.si=3thick, f=(0.8,,0.1,,0.1)}{v4}
\fmffreeze
\fmf{phantom}{v3,v6,v1}
\fmf{dbl_dots_arrow,f=(0.8,,0.1,,0.1),tension=0}{v3,v6}
\fmf{dbl_dots,f=(0.8,,0.1,,0.1),tension=0}{v6,v1}
\end{fmfgraph*}}}
\end{gathered} &
&\begin{gathered} \parbox{30mm} {\resizebox{30mm}{!}{
 \begin{fmfgraph*}(40,18)
 \fmfset{arrow_len}{3mm}
 \fmfset{arrow_ang}{20}
\fmfstraight
\fmfbottom{i1,d1,o1}
\fmftop{i2,d2,o2}
\fmflabel{$n$}{i2}
\fmflabel{$n$}{o2}
\fmflabel{$\bar{n}$}{i1}
\fmflabel{$\bar{n}$}{o1}
\fmf{fermion}{i2,v2}
\fmf{gluon}{v2,v3}
\fmf{gluon}{v3,o2}
\fmf{gluon}{i1,v1}
\fmf{gluon}{v1,v4}
\fmf{fermion}{v4,o1}
\fmffreeze
\fmf{plain}{v2,v3}
\fmf{plain}{v3,o2}
\fmf{plain}{i1,v1}
\fmf{plain}{v1,v4}
\fmf{phantom}{v2,v4}
\fmf{dbl_dots,f=(0.8,,0.1,,0.1),tension=0}{v3,v1}
\fmfv{d.sh=circle, d.fi=full, d.si=3thick, f=(0.8,,0.1,,0.1)}{v2}
\fmfv{d.sh=circle, d.fi=full, d.si=3thick, f=(0.8,,0.1,,0.1)}{v1}
\fmfv{d.sh=circle, d.fi=full, d.si=3thick, f=(0.8,,0.1,,0.1)}{v3}
\fmfv{d.sh=circle, d.fi=full, d.si=3thick, f=(0.8,,0.1,,0.1)}{v4}
\fmffreeze
\fmf{phantom}{v2,v6,v4}
\fmf{dbl_dots_arrow,f=(0.8,,0.1,,0.1),tension=0}{v2,v6}
\fmf{dbl_dots,f=(0.8,,0.1,,0.1),tension=0}{v6,v4}
\end{fmfgraph*}}}
\end{gathered}
\end{align}
%end of figures

%% file: zerobin.tex
\subsubsection{Zero bin subtractions \label{sec:zerobin}}
Finally, we have to consider the overlap between soft, collinear, and Glauber regions. For SCET${}_\text{I}$ problems, with just collinear and ultrasoft modes,the method of regions works without explicit zero bin subtractions in pure dimensional regularization because the soft-collinear zero bin integrals are always scaleless and thus formally vanish. However, even for SCET${}_\text{I}$ problems, one cannot establish the agreement of IR divergences between the full and effective theories without careful consideration of the soft-collinear overlap region. Indeed, factorization requires such zero-bin subtractions to get agreement between full and effective theories at the integrand level~\cite{Manohar:2006nz}. Such subtractions appear, for example, as eikonal jet functions~\cite{css}, or soft-collinear matrix elements~\cite{Feige:2014wja}
In Glauber SCET however, pure dimensional regularization cannot be used, as it does not regulate the rapidity divergences. Thus even the     method-of-regions approach of ignoring the separation of UV and IR and dropping scaleless integrals cannot be used. One must explicitly consider the zero-bin overlap region. 

For quark-gluon scattering, we must subract the zero-bin overlap between the collinear region and the soft and Glauber regions. The full subtracted collinear contribution will then be
\begin{align}\label{eq:zerobin}
    \cI_{\text{pure coll}}=\cI_{\text{coll}}-\cI_{\text{coll}}^{[S]}-\cI_{\text{coll}}^{[G]}+\cI_{\text{coll}}^{[S][G]} \,.
\end{align}
Here, $\cI_{\text{coll}}$ is the collinear contribution, including possible overlaps with the soft and Glauber region, $\cI_{\text{coll}}^{[S]}$ is the collinear-soft zero bin, $\cI_{\text{coll}}^{[G]}$ is the collinear-Glauber zero bin, and $\cI_{\text{coll}}^{[S][G]}$ is the collinear-soft-Glauber zero bin. Each contribution is computed starting from the original collinear graph and then power expanding in the secondary scaling. 

Let us begin with the $n$-collinear quark Wilson line graph $\cI_{WQ,n}$ in Eq.~\eqref{IWQ}. Power expanding in the soft limit, it becomes
\begin{equation}
    i\cI_{WQ,n}^{[S]}
    = g^4  T^a T^b\tilde{\mu}^{2\epsilon}\left(C_F-\frac{C_A}{2}\right)\int \dlipsred{k}{d} \nureg{\nb\cdot k}\frac{2}{[k^2+i0^+]\left[(\nb\cdt p_1)(n\cdt k) +i0^+\right]}\frac{\nb\cdot p_1 }{\nb\cdot k}\ \cM_{T}.
\end{equation}
%Consider now the collinear-soft zero bin $\cI_{WQ,n}^{[S]}$. 
One can perform the $k^+$ integration by contours, picking up the pole from the second propagator. This leaves an integral of the form
\begin{equation}
    \int_{-\infty}^{0}\frac{dk^-}{k^-}\nureg{k^-}=\frac{1}{\eta}-\frac{1}{\eta}=0 \,. % \ & \ I_2&=\int_{-\infty}^{0}dk^-\nureg{k^-}
    \label{4.31}
\end{equation}
Thus with the $\eta$-regulator, $\cI_{WQ,n}^{[S]}=0$. This is a general property of the $\eta$ regulator: the rapidity divergent part of the soft-collinear zero bin will always vanish, as first noted in~\cite{chiu2012formalism,RS}. Note however, that rapidity-finite terms can have non-trivial zero bins, although for $\cI_{WQ,n}$ there are none.

Next, consider the difference between $\cI_{WQ,n}^{[G]}$ and $\cI_{WQ,n}^{[S][G]}$,
\begin{align}
    \cI_{WQ,n}^{[G]}-\cI_{WQ,n}^{[S][G]}
   & =g^2  T^a T^b \tilde{\mu}^{2\epsilon}\left(C_F-\frac{C_A}{2}\right) 
    \int \dlipsred{k}{d} \nureg{\nb\cdot k} \frac{2}{\vp{k}^2} \frac{\nb\cdt p_1}{\nb \cdt k}\nonumber \\* 
    & \times \left[\frac{1}{\nb \cdt p_1\ n\cdt (k +p_1) -(\vp{k}+\vec{p}_{1\perp})^2+i0^+}-\frac{1}{(\nb \cdt p_1)\ (n\cdt k_1) +i0^+}\right]\cM_{T} \,.
\end{align}
Since the $n\cdt k$ poles in the integrands lie on the same side of the contour, this difference vanishes upon contour integration. Combining with Eq.~\eqref{4.31},
\begin{equation}
    \cI_{WQ,n}^{[S]}+\cI_{WQ,n}^{[G]}-\cI_{WQ,n}^{[S][G]}=0\,.
\end{equation} 
A similar analysis shows that the net zero-bin subtraction for the gluon Wilson line graph is also zero,
\begin{equation}
    \cI_{WG,n}^{[S]}+\cI_{WG,n}^{[G]}-\cI_{WG,n}^{[S][G]}=0\,,
\end{equation}
and similarly for the quark $V$ graph,
\begin{equation}
    \cI_{VQ,n}^{[S]}+\cI_{VQ,n}^{[G]}-\cI_{VQ,n}^{[S][G]}=0 \,.
\end{equation}

In contrast to the rest of the collinear graphs, the zero bin subtractions for the gluon V graph $\cI_{VG,n}$ are \emph{non-zero}. Its soft-collinear zero bin  is 
\begin{multline}
    i\cI_{VG,n}^{[S]} =-g^4  T^a T^b\bar{\xi}_{\nb}\left[\slashed{\epsilon}_1-\slashed{k}_{1\perp}\frac{n \cdt \epsilon_{1}}{n \cdt k_1}\right]\frac{1}{\slashed{q}_\perp}C_A\tilde{\mu}^{2\epsilon}   \\*
    \times \int \dlipsred{k}{d} \frac{\mathcal{N}_{VG}^{[S]}}{[k^2+i0^+][(\nb\cdot p_1) (n\cdot k)+i0^+][(k+q_\perp)^2+i0^+]} \xi_{n},
\end{multline}
where the numerator is
\begin{align}
    \mathcal{N}_{VG}^{[S]}
    &=\frac{1}{z}\left(\vec{k}_{\perp}^{2}
    + \vec{k}_{\perp} \cdot \vec{q}_{\perp}\right) \slashed{\epsilon}_{2\perp}
    +\left[\left( \frac{\vec{k}_{\perp}^{2}-\vec{q}_{\perp}^{2}}{2 z}\right)\frac{\slashed{k}_{\perp}}{Q} 
     +\left(\frac{\vec{k}_{\perp}^{2}+\vec{k}_{\perp} \cdot \vec{q}_{\perp}}{z}\right)\frac{\slashed{q}_\perp}{Q} \right] \nb \cdt \epsilon_{2}
     \nonumber \\
   & \hspace{1cm} +\left[\frac{(d-3)}{2}\slashed{k}_{\perp}+\frac{(2+2 z-d z)}{4} \slashed{q}_\perp \right]
    (n\cdt k)(\nb \cdt \epsilon_{2}) \,.
\end{align}
The terms in the first line of the numerator are rapidity divergent and yield zero due to the $\eta$ regulator, in an identical manner to the zero bin subtractions for $\cI_{WQ,n}$. The rest of the terms give a nonzero, rapidity finite integral
\begin{equation}
    \cI_{VG,n}^{[S]}=\frac{g^4}{16\pi^2} T^a T^b C_A \left[\frac{1}{4\epsuv}+\frac{1}{4}\logq+1\right]\left(\cM_{1\nb}+\frac{1}{2}\cM_{q12}\right)\,.
\end{equation}
The rapidity divergent parts of the collinear-Glauber and collinear-Glauber-soft zero bins for $\cI_{VG,n}$ vanish, and the rapidity-finite parts are power suppressed,
\begin{equation}
 \cI_{VG,n}^{[G]}-\cI_{VG,n}^{[S][G]}=0 \,.
\end{equation}
The total zero bin contribution for the gluon V graph is
\begin{equation}
    \cI_{VG,n}^{[S]}+\cI_{VG,n}^{[G]}-\cI_{VG,n}^{[S][G]}=  \cI_{VG,n}^{[S]} \not=0 \,.
\end{equation}
A simplified calculation that illustrates how the zero-bin for the gluon V graph can be nonvanishing is provided in Appendix~\ref{app:zerobintoy}. 
This is exactly what is needed to restore the Ward identity when subtracted from the collinear contribution in Eq.~\eqref{Incoll}.

One must also consider soft-Glauber zero bin contributions. For a general soft graph, the zero bin subtractions take the form 
\begin{equation}\label{eq:softzero}
    \cI_\text{pure soft}=\cI_\text{soft}-\cI_\text{soft}^{[G]}\ ,
\end{equation}
where $\cI_\text{soft}^{[G]}$ is the soft-Glauber zero bin. In~\cite{RS} it was argued that whether the rapidity-divergent parts of soft-Glauber zero bin integrals vanish depends on the convention one chooses for the $i0^\pm$ prescription in the soft Wilson lines. Different choices shift a contribution between $\cI_\text{soft}$ and $\cI_\text{soft}^{[G]}$, but their difference is independent of convention. We illustrate this for the flower graph in Appendix~\ref{app:flowerzero}.
In the natural convention where the eikonal propagators inherit $+ i 0^+$ terms from the full-theory graphs, the rapidity-divergent parts of the zero bin vanish. 
One must also compute the rapidity-finite parts of the soft-Glauber zero bin contribution. As with the gluon V graph soft-Glauber contribution, the rapidity-finite part of the soft graph zero bins are all power suppressed or vanish.
Hence, all the soft graphs do not receive any soft-Glauber zero bin subtractions:
\begin{equation}
    \cI_\text{eye}^{[G]} = \cI_\text{flower}^{[G]} =0 \,.
\end{equation}

In summary, almost all of the zero bin subtractions give zero. The only non-vanishing contribution comes from the collinear-soft zero bin of the gluon V graph, so the total zero bin contribution is
\begin{equation}
\label{Izerobin}
    \cI_{\text{zero bin}}= -\cI_{VG,n}^{[S]} = -\frac{g^4}{16\pi^2} T^a T^b C_A \left[\frac{1}{4\epsuv}+\frac{1}{4}\logq+1\right]\left(\cM_{1\nb}+\frac{1}{2}\cM_{q12}\right)\,.
\end{equation}

%% file: matching.tex
\subsection{Final result}

Adding up the results from Eqs.~\eqref{Incoll}, \eqref{Isoft}, \eqref{Iboxes} and \eqref{Izerobin}, and including the $\nb$-collinear contribution as well, we get
\begin{align}
&\cI_{\text{SCET}}^1 = \cI_{\text{coll},n} + \cI_{\text{zerobin},n} + \cI_{\text{coll},\nb}  + \cI_{\text{zerobin},\nb} + \cI_{\text{soft}} + \cI_{\text{boxes}}
\nonumber \\
    &=\frac{g^4}{16\pi^2}C_F\left[\frac{1}{\epsuv}-\frac{2}{\epsir^2}-\frac{4+2\ln{\frac{\mu^2}{Q^2}}}{\epsir}-2\ln{\frac{\mu^2}{Q^2}}\logq+\ln^2{\frac{\mu^2}{-t}}-3\logq+\frac{7\pi^2}{6}-9 \right]\cM_{T} (T^a T^b)_{ij} \nonumber \\
    &+\frac{g^4}{16\pi^2}C_A\left[\frac{2}{\epsuv}-\frac{2}{\epsir^2}-\frac{2+2\logq}{\epsir}-\ln^2{\frac{\mu^2}{-t}}+\frac{\pi^2}{6}+2\right]\cM_{T}(T^a T^b)_{ij}\nonumber \\
    &-\frac{g^4}{16\pi^2}(C_A-C_F)(2\cM_{L1}+2\cM_{L2})(T^a T^b)_{ij} +\frac{ g^4}{16\pi^2} \delta^{ab}\delta_{ij} \left[\frac{1}{\epsir}+\logq\right](i\pi) .
\end{align}
Comparing to the leading-power expansion  of the full QCD amplitude in Eq.~\eqref{IQCD1} we find perfect agreement. Thus there are no matching corrections at 1-loop:
\begin{equation}
  \begin{boxed}{
      C_T = 1  \,, \quad
      C_{L1} = C_{L2} = C_{L3} = C_{L4} = 0\,.
       }
       \end{boxed} 
    \label{wilsoncoeff}
\end{equation}
Although we cannot rule out contributions to these Wilson coefficients at order $\alpha_s^2$ and above, there is good reason to suspect such contributions will also vanish.

% Comparing to the leading-power expansion  of the full QCD amplitude in Eq.~\eqref{IQCD1} we find nearly perfect agreement. All the UV-divergent, IR-divergent and rapidity divergent terms agree, all the logarithms agree, all the new tensor structure contributions ($\cM_{L1}$ and $\cM_{L2}$) agree, the Glauber phase contribution agrees and the entire $C_A$ contribution agrees. The only difference is in a finite part proportional to $C_F\cM_T$:
% \begin{equation}
% \begin{boxed}{
%     \cI_{\text{QCD}}^1 - \cI_{\text{SCET}}^1 = 2\alpha_s^2 C_F  \cM_T \,.
%     }
% \end{boxed} 
% \label{4.42}
% \end{equation}
% This is the main result of our paper. It implies that there is a finite one-loop matching correction to the coefficient of the quark Glauber operator present at tree-level, and the other operators in Eq.~\eqref{OFG} do not appear at one-loop,
% \begin{equation}
%  \begin{boxed}{
%       C_T = 1 + \frac{\alpha_s}{2\pi} C_F \,, \quad
%       C_{L1} = C_{L2} = C_{L3} = C_{L4} = 0\,.
%       }
%       \end{boxed} 
%     \label{wilsoncoeff}
% \end{equation}
% Eq.~\eqref{4.42} and Eq.~\eqref{wilsoncoeff} also hold for QED with the replacement $C_F \to 1$.

%% file: conclusions.tex
\section{Conclusions}

In this paper, we have studied quark-gluon backscattering in the Regge limit, where the outgoing quark is close to the direction of the incoming gluon. We work to leading power in $\lambda = {|q_\perp|}/{Q}$, with $q_\perp$ the transverse momentum of the outgoing quark relative to the incoming gluon and $Q$ its energy. We computed the one-loop amplitude in QCD in this limit as well as the one-loop contribution to the amplitude using Glauber-SCET. We find that the same operators $\cO_T$, $\cO_{sn},\cO_{s\nb}$ and $\cO_{GG0}$ required for tree-level matching also reproduce all of the UV divergences, IR divergences, large logarithms and all the finite parts of QCD.% There is, however, a finite matching correction needed, although even this is remarkably simple. The fermionic Glauber operator at one-loop is
%\begin{equation}
%    \cO_{FG} =  \left[1 + \frac{\alpha_s}{2\pi} C_F\right] \cO_T\,.
%    \label{OFGform}
%\end{equation}
%With this form, even the finite parts of the QCD ampliude are reproduced exactly, to leading power in the Regge limit. 

Our result in Eq.~\eqref{wilsoncoeff} is very surprising. In principle, five different operators could have been needed at one-loop. Indeed, we did find that the Dirac tensor structures corresponding to $\cO_{L1}$ and $\cO_{L2}$ do appear at one-loop. However, loop corrections from insertions of $\cO_T$ generate these tensor structures, so that $\cO_{L1}$ and $\cO_{L2}$ are not needed in the Glauber-SCET Lagrangian at one-loop.
We find that the only operator needed is $\cO_T$, the same one needed for tree-level matching, but even for this operator, the one-loop matching correction vanishes.

That there is no one-loop correction can be anticipated from general arguments, such as those in~\cite{RS}. Indeed, for quark-quark forward scattering in the Regge limit, Stewart and Rothstein found that not only was the leading power behavior of QCD reproduced at one-loop by Glauber-SCET, but that even the finite correction to the gluonic Glauber operator vanished. General arguments given in~\cite{RS} forbid dependence on the hard scale, so that terms like $\ln({p_1\cdot p_2}/{\mu^2})$ should not appear in the Wilson coefficient. This can explain the absence of logarithms in the matching, both for gluonic Glauber exchange and fermionic exchange, but does not forbid a correction which is a function of $\alpha_s$. We find that just as the Glauber gluonic operator, the Glauber fermionic operator in SCET reproduces the entire leading power Regge behaviour in QCD, including the finite terms, so there is no $\alpha_s$ correction.

 The one-loop matching for the Glauber gluon case enters the analysis of the two-loop gluonic Regge trajectory~\cite{regge}, and the computation is simplified because the matching vanishes. The one-loop matching for the Glauber quark case computed in this paper (which also vanishes) can be similarly used to compute the two-loop quark Regge trajectory.\footnote{We thank Ian Moult for discussions on~\cite{regge} prior to publication.} The calculational method requires the eye graph, to order $\epsilon$, which is given in Appendix~\ref{sec:eye}.

\subsection*{Acknowledgements}

We would like to thank I.~Moult, D.~Neill, M.~Solon, I.~Stewart, W.~Waalewijn, K.~Yan, and H-X. Zhu for discussions. MDS and AB are supported in part by the  U.S. Department of Energy under contract DE-SC0013607. AM is supported in part by the U.S. Department of Energy under contract DE-SC0009919.

%% file: Wilson.tex
\section{Glauber SCET expansion}
In this appendix we discuss some subtleties in expanding the Glauber SCET operators and give the Feynman rules explicitly.

\subsection{Wilson lines \label{sec:Wilson}}
For $n$-collinear fields, the Wilson lines $W_n$ represent $n$-collinear radiation from everything else in the event, i.e.\ from other particles in the collision which are \emph{not} in the $n$ direction.
This radiation can come from incoming or outgoing particles, but the $n$-collinear field is only sensitive to the net effect which can be treated as coming from a coherent source of radiation from a single outgoing particle propagating in the $\nb$ direction, and with color charge conjugate to the $n$-collinear field. The Wilson line takes the form\footnote{A reminder that our sign convention for the covariant derivative is $D_\mu=\partial_\mu - i g A_\mu$.} 
\begin{equation}
W_{n}^\dag(x)=P\left\{ \exp\left[ig T^a \int_0^{\infty} ds\ \nb\cdt A_n^a(x^\nu+\nb^\nu s) e^{-0^+ s} \right]\right\}
\end{equation}
where $P$ denotes path-ordering. Under a gauge transformation $W_n^\dagger$ transforms as
\begin{align}
  W_{n}^\dag(x) &\to U(\infty) W_{n}^\dag(x) U^\dag (x)  \,.
\end{align}
The $\exp^{-0^+ s}$ factor is required to make the Wilson line convergent at infinity. It generates the $i0^+$ factors in the propagator denominators. Note that $W_{n}^\dag(x)$ contains $n$-collinear gauge fields, and the integral is in the $\nb$ direction.
For $\nb$-collinear fields one uses Wilson lines pointing in the $n$ direction.
\begin{equation}
W^\dag_{\nb}(x)=P\left\{ \exp\left[ig T^a \int_0^{\infty} ds\ n\cdt A_{\nb}^a(x^\nu+n^\nu s) e^{-0^+ s} \right]\right\} \,.
\end{equation}

Soft Wilson lines represent soft radiation from the collinear particles, and point in the direction of those particles (in contrast to collinear Wilson lines which represent radiation from everything else {\it{except}} the collinear particle of interest). For the soft operators, we need Wilson lines pointing both in the $n$ and $\nb$ direction. These Wilson lines have identical structure to the collinear Wilson lines, but comprising soft rather than collinear gluon fields, and conventionally indexed with the direction of the Wilson line path rather than the direction backwards to the Wilson line path. Explicitly, for outgoing radiation
\begin{align}
S^\dag_{n}(x)&=P\left\{ \exp\left[ig T^a \int_0^{\infty} ds\ n\cdt A_S^a(x^\nu+n^\nu s) e^{-0^+ s} \right]\right\}\,,  \\
S^\dag_{\nb}(x)&=P\left\{ \exp\left[ig T^a \int_0^{\infty} ds\ \nb\cdt A_S^a(x^\nu+\nb^\nu s) e^{-0^+ s} \right]\right\}\,,
\end{align}
which transform as
\begin{align}
   S^\dag_{n}(x)&\to U(\infty) S^\dag_{n}(x) U^\dag(x)\,, & 
\  S^\dag_{\nb}(x)&\to U(\infty) S^\dag_{\nb}(x) U^\dag(x) \,.
\end{align}
under gauge transformations.
Since forward scattering involves incoming as well as outgoing particles, we should also consider soft operators using Wilson lines representing incoming radiation:
\begin{align}
\overline{S}_{n}(x)&=P\left\{ \exp\left[ig T^a \int_{-\infty}^0 ds\ n\cdt A_S^a(x^\nu+n^\nu s) e^{0^+ s} \right]\right\} \,,  \\
\overline{S}_{\nb}(x)&=P\left\{ \exp\left[ig T^a \int_0^{\infty} ds\ \nb\cdt A_S^a(x^\nu+\nb^\nu s) e^{0^+ s} \right]\right\} \,,
\end{align}
which transform as
\begin{align}
   \overline{S}^\dag_{n}(x)&\to U(-\infty) \overline{S}^\dag_{n}(x) U^\dag(x)\,, & 
\  \overline{S}^\dag_{\nb}(x)&\to U(-\infty) \overline{S}^\dag_{\nb}(x) U^\dag(x) \,.
\end{align}
under gauge transformations.
The difference between $S$ and $\overline{S}$ is $+i0^+ \to -i 0^+$ in the eikonal propagators. One uses $\overline{S}$ or $S$ depending on whether the QCD field $\psi$ represents an incoming quark or an outgoing antiquark.

%% file: feynmanrules.tex
\subsection{Collinear gluon operators}

SCET $n$-collinear gluons are created and annihilated by the gauge invariant operator $\mathcal{B}_{n\perp}^\mu$, which is defined by
\begin{align}
   \mathcal{B}_{n\perp}^\mu &= \left[ \frac{i {\bar n}_\nu}{\bar n \cdot \mathcal{P}} W^\dagger_{n} G_n^{\nu \mu_\perp} W_{n} \right]\,,
\end{align}
where $G_n^{\nu \mu}$ is the gluon field-strength tensor for $n$-collinear gluons, $G_n^{\nu \mu_\perp}$
projects the $\mu$ index to the $\perp$ directions, and the label operator $\mathcal{P}$ is the total label momentum of $W^\dagger_{n} G^{\nu \mu_\perp} W_{n}$. The square brackets are a reminder that $\mathcal{P}$ picks out the label momentum of the fields inside the brackets. The one-gluon matrix element of $\mathcal{B}_{n\perp}^\mu$ for a gluon with polarization $\epsilon$ and incoming momentum $k$ is
\begin{align}
    \left(\epsilon^\mu - \frac{\bar n \cdot \epsilon}{\bar n \cdot k} k^\mu \right) T^a\,,
    \label{B.121}
\end{align}
so that $\mathcal{B}_{n\perp}^\mu$ acts like a gauge-invariant version of $A_\mu$.

There is an alternate definition of $\mathcal{B}_{n\perp}^\mu$, $g \mathcal{B}_{n\perp}^\mu= W^\dagger_{n}i D^\mu_\perp W_{n}$ which is equivalent to Eq.~\eqref{B.121}. Equation~\eqref{B.121} is more convenient for our purposes because it simplifies the computation of the gluon matrix elements.

The $n$-collinear operator we need in the Glauber theory is
\begin{align}
    \mathcal{B}^\mu_{n \perp} \chi_n = 
    \left[ \frac{i {\bar n}_\nu}{\bar n \cdot \mathcal{P}} W^\dagger_{n} G_n^{\nu \mu_\perp} W_{n} \right] \left[W^\dagger_{n} \xi_n \right]\,,
\end{align}
where the label momentum $\mathcal{P}$ only acts on the fields within the first square brackets.
The matrix element of $ \mathcal{B}^\mu_{n \perp} \chi_n$ with an incoming fermion and one incoming gluon is
\begin{align}
    \left(\epsilon^\mu - \frac{\bar n \cdot \epsilon}{\bar n \cdot k} k^\mu \right) T^a \xi\,.
\end{align}
For one incoming fermion and two incoming gluons, the matrix element is
\begin{align}
   g \frac{\bar n \cdot \epsilon_2}{ \bar n \cdot k_2} \epsilon_{1\nu} T^{a_2} T^{a_1} \xi + g   \frac{\bar n \cdot \epsilon_1}{ \bar n \cdot k_1}  \epsilon_{2\nu} T^{a_1} T^{a_2} \xi
  - \frac{g ( \bar n \cdot \epsilon_1) (\bar n \cdot \epsilon_2) }{\bar n \cdot (k_1+k_2)} \left(k_{1\nu}+k_{2\nu} \right) \left[ \frac{1}{ \bar n \cdot k_2}  T^{a_2} T^{a_1}  +  \frac{1}{ \bar n \cdot k_1}   T^{a_1} T^{a_2} \right] \xi 
\end{align}
where $\xi$ is the fermion spinor, and the gluons have momenta $k_i$ and color $a_i$. These results are used in the next section to get the Glauber-SCET Feynman rules.

\subsection{Feynman Rules}
In this Appendix we summarize the Feynman rules of Glauber-SCET that we use for the calculations in this paper. These are mostly well known, and we simply copy them from~\cite{RS,MSSV,bauerBXs,bauer2001effective,bauer2001invariant,bauer2002hard,bauer2002power,bauer2002soft}. 
The only new vertex is the 2 collinear gluon emission off the Glauber operator in Eq.~\eqref{twocollFR}. This was not given in~\cite{MSSV}, and moreover, we disagree with Eq. (A.5) of~\cite{MSSV} which gives the second order terms in the expansion of the collinear operator.
We find the second order terms in the expansion of $\cB_{n\perp}^\mu$ are
\begin{align}
&- g  \left[T^{a_1},T^{a_2}\right]  \left[ \frac{(\bar n \cdot \epsilon_1)(\bar n \cdot \epsilon_2)}{\bar n \cdot (k_1+k_2)} \left( \frac{k_{2\nu}}{\bar n \cdot k_1} - \frac{k_{1\nu}}{\bar n \cdot k_2} \right)
+ \frac{\epsilon_{1\nu} (\bar n \cdot  \epsilon_2)}{\bar n \cdot k_2}  -  \frac{\epsilon_{2\nu} (\bar n \cdot  \epsilon_1)}{\bar n \cdot k_1} \right]\,.
\label{6.11}
\end{align}

The Glauber SCET Feynman rules used in this paper are as follows.
The $n$-collinear Glauber vertex with one collinear gluon emission is
\begin{equation}
     \begin{gathered} \parbox{30mm} {\resizebox{30mm}{!}{
  \begin{fmfgraph*}(35,20)
    \fmfset{arrow_len}{3mm}
    \fmfset{arrow_ang}{20}
    \fmfstraight
    \fmfbottom{i1,d1,o1}
    \fmftop{i2,d2,o2}
    \fmflabel{$\mu,a$}{o2}
    \fmf{phantom}{i1,v1}
    \fmf{phantom}{v1,o1}
    \fmf{fermion}{i2,v2}
    \fmf{gluon}{v2,o2}
    \fmffreeze
    \fmf{plain}{v2,o2}
    \fmf{dbl_dots_arrow,f=(0.8,,0.1,,0.1)}{v2,v1}
    \fmfv{d.sh=circle, d.fi=full, d.si=2thick, f=(0.8,,0.1,,0.1)}{v2}
    \fmfcmd{style_def karrow expr p = drawarrow subpath (1/4, 3/4) of p shifted 12 down
        withpen pencircle scaled 0.4; label.top(btex $k$ etex, point 0.5 of p
        shifted 25 down); enddef;}
    \fmf{karrow}{v2,o2}
  \end{fmfgraph*}}}
\end{gathered}
    =T^a\left(\gamma_\perp^\mu-\slashed{k}_\perp \frac{\nb^\mu}{\nb\cdot k}\right) \,.
\end{equation}
\\
With two collinear gluons it is 
\begin{equation}
    \begin{gathered} \parbox{30mm} {\resizebox{30mm}{!}{
  \begin{fmfgraph*}(35,20)
    \fmfstraight
    \fmfleft{i1,i2,i3}
    \fmfright{o1,o2,o3}
    \fmf{phantom}{i1,v1}
    \fmf{phantom}{v1,o1}
    \fmf{fermion}{i3,v3}
    \fmf{gluon}{v3,o3}
    \fmffreeze
    \fmf{dbl_dots_arrow,f=(0.8,,0.1,,0.1)}{v3,v1}
    \fmf{gluon}{v3,o2}
    \fmf{plain}{v3,o3}
    \fmf{plain}{v3,o2}
    \fmflabel{$\mu,a$}{o3}
    \fmflabel{$\nu,b$}{o2}
    \fmfv{d.sh=circle, d.fi=full, d.si=2thick, f=(0.8,,0.1,,0.1)}{v3}
    \fmfcmd{style_def karrow expr p = drawarrow subpath (1/4, 3/4) of p shifted 10 up
        withpen pencircle scaled 0.4; label.top(btex $k_1$ etex, point 0.5 of p
        shifted 12 up); enddef;}
    \fmf{karrow}{o3,v3}
    \fmfcmd{style_def kparrow expr p = drawarrow subpath (1/4, 3/4) of p shifted 14 down
        withpen pencircle scaled 0.4; label.top(btex $k_2$ etex, point 0.5 of p
        shifted 28 down); enddef;}
    \fmf{kparrow}{o2,v3}
    \fmfset{arrow_len}{3mm}
    \fmfset{arrow_ang}{20}
  \end{fmfgraph*}}}
\end{gathered}
\hspace{15mm}
=\hspace{2mm}
\begin{aligned}
    & -g \left(\slashed{k}_{1 \perp}+\slashed{k}_{2 \perp}\right)\left(\frac{T^{a} T^{b}}{\nb \cdot k_{1}}    +\frac{T^{b} T^{a}}{\nb \cdot k_{2}}\right) \frac{\nb^\mu \nb^\nu}{\nb \cdot k_{1}+\nb \cdot k_{2}}\\
   & \hspace{40mm} + g T^{a} T^{b} \frac{\nb^\mu}{\nb \cdot k_{1}} \gamma_{\perp}^\nu+ g T^{b} T^{a} \frac{\nb^\nu}{\nb \cdot k_{2}} \gamma_{\perp}^\mu \,.
    \end{aligned}
    \label{twocollFR}
\end{equation}
\\[3mm]
The vertex for a single soft emission is~\cite{MSSV}
\begin{equation}
\begin{gathered} \parbox{30mm} {\resizebox{13mm}{!}{
  \begin{fmfgraph*}(15,20)
    \fmfset{arrow_len}{3mm}
    \fmfset{arrow_ang}{20}
    \fmfstraight
    \fmfleft{i1,i2,i3}
    \fmfright{o1}
    \fmf{dbl_dots_arrow,f=(0.8,,0.1,,0.1)}{i2,i1}
    \fmf{dbl_dots_arrow,f=(0.8,,0.1,,0.1)}{i3,i2}
    \fmf{gluon,f=(0.0,,0.5,,0.0)}{i2,o1}
    \fmflabel{$\mu,a$}{o1}
    \fmfcmd{style_def karrow expr p = drawarrow subpath (1/4, 3/4) of p shifted 12 down
        withpen pencircle scaled 0.4; label.top(btex $k$ etex, point 0.5 of p
        shifted 28 down); enddef;}
    \fmf{karrow}{i2,o1}
    \fmfcmd{style_def qarrow expr p = drawarrow subpath (1/4, 3/4) of p shifted 12 left
        withpen pencircle scaled 0.4; label.top(btex $q_\perp$ etex, point 0.5 of p
        shifted 20 left); enddef;}
    \fmf{qarrow}{i3,i2}
    \fmfcmd{style_def qparrow expr p = drawarrow subpath (1/4, 3/4) of p shifted 12 left
        withpen pencircle scaled 0.4; label.top(btex $q_\perp^\prime$ etex, point 0.5 of p
        shifted 20 left); enddef;}
    \fmf{qparrow}{i2,i1}
    \fmfv{d.sh=circle, d.fi=full, d.si=2thick, f=(0.8,,0.1,,0.1)}{i2}
  \end{fmfgraph*}}}
\end{gathered}
\hspace{-5mm}
= g^3 T^a \left(\gamma^\mu_\perp +\slashed{q}_\perp^\prime \frac{n^\mu}{n\cdt k}-\slashed{q}_\perp \frac{\nb^\mu}{\nb\cdt k}\right)\,.
\end{equation}
For two soft emissions:
\begin{equation}
\begin{gathered} \parbox{30mm} {\resizebox{13mm}{!}{
  \begin{fmfgraph*}(15,20)
    \fmfset{arrow_len}{3mm}
    \fmfset{arrow_ang}{20}
    \fmfstraight
    \fmfleft{i1,i2,i3}
    \fmfright{o3,o1,o4,o2,o5}
    \fmf{dbl_dots_arrow,f=(0.8,,0.1,,0.1)}{i2,i1}
    \fmf{dbl_dots_arrow,f=(0.8,,0.1,,0.1)}{i3,i2}
    \fmf{gluon,f=(0.0,,0.5,,0.0)}{i2,o1}
    \fmf{gluon,f=(0.0,,0.5,,0.0)}{i2,o2}
    \fmflabel{$\mu,a$}{o1}
    \fmflabel{$\nu,b$}{o2}
    \fmffreeze
    \fmfcmd{style_def karrow expr p = drawarrow subpath (1/4, 3/4) of p shifted 12 down
        withpen pencircle scaled 0.4; label.top(btex $k_1$ etex, point 0.5 of p
        shifted 28 down); enddef;}
    \fmf{karrow}{i2,o1}
    \fmfcmd{style_def ksarrow expr p = drawarrow subpath (1/4, 3/4) of p shifted 10 up
        withpen pencircle scaled 0.4; label.top(btex $k_2$ etex, point 0.5 of p
        shifted 15 up); enddef;}
    \fmf{ksarrow}{i2,o2}
    \fmfcmd{style_def qarrow expr p = drawarrow subpath (1/4, 3/4) of p shifted 12 left
        withpen pencircle scaled 0.4; label.top(btex $q_\perp$ etex, point 0.5 of p
        shifted 20 left); enddef;}
    \fmf{qarrow}{i3,i2}
    \fmfcmd{style_def qparrow expr p = drawarrow subpath (1/4, 3/4) of p shifted 12 left
        withpen pencircle scaled 0.4; label.top(btex $q_\perp^\prime$ etex, point 0.5 of p
        shifted 20 left); enddef;}
    \fmf{qparrow}{i2,i1}
    \fmfv{d.sh=circle, d.fi=full, d.si=2thick, f=(0.8,,0.1,,0.1)}{i2}
  \end{fmfgraph*}}}
\end{gathered}
\hspace{-10mm}
=
\hspace{3mm}
\begin{aligned}
& g^4  T^a T^b
\Big(
\frac{n^\nu\gamma^\mu_\perp}{n\cdt k_2}
-\frac{\nb^\nu\gamma^\mu_\perp}{\nb\cdt k_1}
+\frac{\slashed{q}_\perp^\prime \nb^\mu\nb^\nu}{2 \nb\cdt (k_1+k_2) \nb\cdt k_1}
-\frac{\slashed{q}_\perp n^\mu n^\nu}{2 n\cdt (k_1+k_2) n\cdt k_1}\\
&\hspace{50mm}
 -\frac{(\slashed{q}_\perp+\slashed{k}_{1\perp})\nb^\mu n^\nu}{\nb\cdt k_1 n\cdt k_2 }\Big)
+(k_1,\mu,a)\leftrightarrow (k_2,\nu,b)\ .
\end{aligned}
\end{equation}
The soft-quark/soft-gluon Glauber vertex (from $\cO_{Sn}$) is
\begin{equation}
    \begin{gathered} \parbox{30mm} {\resizebox{30mm}{!}{
  \begin{fmfgraph*}(35,20)
    \fmfstraight
    \fmfleft{i1,i2}
    \fmfright{o1,o2}
    \fmf{phantom}{i1,v1}
    \fmf{phantom}{v1,o1}
    \fmf{gluon,f=(0,,0.5,,0.0)}{v2,o2}
    \fmf{fermion,f=(0,,0.5,,0.0)}{i2,v2}
    \fmflabel{$\mu,a$}{o2}
    \fmffreeze
    \fmf{dbl_dots_arrow,f=(0.8,,0.1,,0.1)}{v2,v1}
    \fmfv{d.sh=circle, d.fi=full, d.si=2thick, f=(0.8,,0.1,,0.1)}{v2}
    \fmfcmd{style_def garrow expr p = drawarrow subpath (1/4, 3/4) of p shifted 12 down
        withpen pencircle scaled 0.4; label.top(btex $k$ etex, point 0.5 of p
        shifted 24 down); enddef;}
    \fmf{garrow}{v2,o2}
    \fmfset{arrow_len}{3mm}
    \fmfset{arrow_ang}{20}
  \end{fmfgraph*}}}
\end{gathered}=-g^2 T^a\left(\gamma_{\perp}^\mu-\slashed{k}_\perp \frac{\nb^\mu}{\nb\cdt k}\right)u_{s}(p)\,.
\end{equation}
\\
\\
The 3-point collinear-quark/collinear-gluon vertex in SCET is
\\[5mm]
\begin{equation}
    \begin{gathered} \parbox{30mm} {\resizebox{30mm}{!}{
  \begin{fmfgraph*}(35,15)
    \fmfstraight
    \fmfleft{i1,i2}
    \fmfright{o1,o2}
    \fmf{fermion}{i1,v1}
    \fmf{phantom}{v2,o2}
    \fmf{phantom}{i2,v2}
    \fmf{fermion}{v1,o1}
    \fmffreeze
    \fmf{gluon}{v1,v2}
    \fmf{plain}{v1,v2}
    \fmflabel{$a,\mu$}{v2}
    \fmfcmd{style_def parrow expr p = drawarrow subpath (1/4, 3/4) of p shifted 12 down
        withpen pencircle scaled 0.4; label.top(btex $p$ etex, point 0.5 of p
        shifted 24 down); enddef;}
    \fmfcmd{style_def pnarrow expr p = drawarrow subpath (1/4, 3/4) of p shifted 12 down
        withpen pencircle scaled 0.4; label.top(btex $p^\prime$ etex, point 0.5 of p
        shifted 24 down); enddef;}
    \fmf{parrow}{i1,v1}
    \fmf{pnarrow}{v1,o1}
    \fmfset{arrow_len}{3mm}
    \fmfset{arrow_ang}{20}
  \end{fmfgraph*}}}
\end{gathered}=ig T^a\left[n^\mu +\frac{\gamma^\mu_\perp \slashed{p}_\perp}{\nb\cdt p}+\frac{\slashed{p}_\perp^\prime \gamma^\nu_\perp}{\nb\cdt p^\prime} -\frac{\slashed{p}_\perp^\prime \slashed{p}_\perp}{(\nb\cdt p)(\nb\cdt p^\prime)}\nb^\mu\right]
\frac{\slashed{\nb}}{2} \,.
\end{equation}
\\[5mm]
The 4 point collinear-quark/collinear-gluon vertex in SCET is 
\\[5mm]
\begin{align}
    \begin{gathered} \parbox{30mm} {\resizebox{30mm}{!}{
  \begin{fmfgraph*}(35,15)
    \fmfstraight
    \fmfleft{i1,i2}
    \fmfright{o1,o2}
    \fmf{fermion}{i1,v1}
    \fmf{phantom}{v2,o2}
    \fmf{phantom}{i2,v2}
    \fmf{fermion}{v1,o1}
    \fmffreeze
    \fmf{gluon}{v1,i2}
    \fmf{plain}{v1,i2}
    \fmflabel{$a,\mu$}{i2}
    \fmf{gluon}{v1,o2}
    \fmf{plain}{v1,o2}
    \fmflabel{$b,\nu$}{o2}
    \fmfcmd{style_def parrow expr p = drawarrow subpath (1/4, 3/4) of p shifted 12 down
        withpen pencircle scaled 0.4; label.top(btex $p$ etex, point 0.5 of p
        shifted 24 down); enddef;}
    \fmfcmd{style_def pnarrow expr p = drawarrow subpath (1/4, 3/4) of p shifted 12 down
        withpen pencircle scaled 0.4; label.top(btex $p^\prime$ etex, point 0.5 of p
        shifted 24 down); enddef;}
    \fmfcmd{style_def karrow expr p = drawarrow subpath (1/4, 3/4) of p shifted 18 up
        withpen pencircle scaled 0.4; label.top(btex $k$ etex, point 0.5 of p
        shifted 24 up); enddef;}
    \fmf{parrow}{i1,v1}
    \fmf{pnarrow}{v1,o1}
    \fmf{karrow}{v1,o2}
    \fmfset{arrow_len}{3mm}
    \fmfset{arrow_ang}{20}
  \end{fmfgraph*}}}
\end{gathered}&=ig^2 \frac{T^a T^b}{\nb\cdt(p-k)}\left[\gamma_{\perp}^\mu \gamma^\nu_\perp -\frac{\gamma^\mu_\perp \slashed{p}_\perp}{\nb\cdt p}\nb^\nu-\frac{\slashed{p}_\perp^\prime \gamma^\nu_\perp}{\nb\cdt p^\prime}\nb^{\mu} +\frac{\slashed{p}_\perp^\prime \slashed{p}_\perp}{(\nb\cdt p)(\nb\cdt p^\prime)}\nb^\mu\nb^\nu\right]
\frac{\slashed{\nb}}{2}\nonumber\\
&+ig^2 \frac{T^b T^a}{\nb\cdt(k+p^\prime)}\left[\gamma_{\perp}^\nu \gamma^\mu_\perp -\frac{\gamma^\nu_\perp \slashed{p}_\perp}{\nb\cdt p}\nb^\mu-\frac{\slashed{p}_\perp^\prime \gamma^\mu_\perp}{\nb\cdt p^\prime}\nb^{\nu} +\frac{\slashed{p}_\perp^\prime \slashed{p}_\perp}{(\nb\cdt p)(\nb\cdt p^\prime)}\nb^\nu\nb^\mu\right]
\frac{\slashed{\nb}}{2}
\end{align}
\\[5mm]
The 3-point collinear gluon vertex is the same as that of QCD,
\\[5mm]
\begin{align}
    \begin{gathered} \parbox{30mm} {\resizebox{30mm}{!}{
  \begin{fmfgraph*}(35,20)
    \fmfleft{i1}
    \fmfright{o1,o2}
    \fmf{gluon,tension=1.7}{i1,v1}
    \fmf{gluon}{v1,o1}
    \fmf{gluon}{v1,o2}
    \fmffreeze
    \fmf{plain}{i1,v1}
    \fmf{plain}{v1,o1}
    \fmf{plain}{v1,o2}
    \fmflabel{$\nu,b$}{i1}
    \fmflabel{$\rho,c$}{o1}
    \fmflabel{$\mu,a$}{o2}
    \fmfcmd{style_def parrow expr p = drawarrow subpath (1/4, 3/4) of p shifted 12 down
        withpen pencircle scaled 0.4; label.top(btex $p$ etex, point 0.5 of p
        shifted 24 down); enddef;}
    \fmfcmd{style_def qarrow expr p = drawarrow subpath (1/4, 3/4) of p shifted 18 down
        withpen pencircle scaled 0.4; label.top(btex $q$ etex, point 0.5 of p
        shifted 32 down); enddef;}
    \fmfcmd{style_def karrow expr p = drawarrow subpath (1/4, 3/4) of p shifted 16 up
        withpen pencircle scaled 0.4; label.top(btex $k$ etex, point 0.5 of p
        shifted 20 up); enddef;}
    \fmf{parrow}{i1,v1}
    \fmf{qarrow}{o1,v1}
    \fmf{karrow}{o2,v1}
    \fmfset{arrow_len}{3mm}
    \fmfset{arrow_ang}{20}
  \end{fmfgraph*}}}
\end{gathered}&=gf^{abc}\left[g^{\mu\nu}(k-p)^\rho+g^{\nu\rho}(p-q)^\mu +g^{\rho\mu}(q-k)^\nu\right]\,.
\end{align}
\\[5mm]
Finally, we also need the leading power collinear quark propagator, 
\begin{align}
    \begin{gathered}  {\resizebox{20mm}{!}{
  \begin{fmfgraph*}(20,10)
    \fmfleft{i1}
    \fmfright{o1}
    \fmf{fermion,label=$p$}{i1,o1}
    \fmfset{arrow_len}{3mm}
    \fmfset{arrow_ang}{20}
  \end{fmfgraph*}}}
\end{gathered}=i\frac{\nb\cdt p}{p^2+i0^+}\frac{\slashed{n}}{2}\,.
\end{align}

%% file: zerobintoy.tex
\section{Zero bin toy calculation \label{app:zerobintoy}}
In this appendix, we provide some illustrative calculations relevant for the zero-bin subtractions needed in this paper.

\subsection{Collinear-Soft zero bin \label{app:collinearsoft}}
Consider the collinear integral
\begin{equation}
    \tilde{I}_\text{toy}=\tilde{\mu}^{2\epsilon}\int\dlipsred{k}{d}\frac{(n\cdot k)\ (\nb\cdot p_1)}{[k^2+i0^+][(k+p_1)^2+i0^+][(k+q_\perp)^2+i0^+]}\,.
    \label{toy}
\end{equation}
This is a standard one-loop integral with quadratic denominators. It evaluates to
\begin{equation}
    \tilde{I}_\text{toy}=\frac{i}{16\pi^2}\left[\frac{1}{4\epsir^2}+\frac{2+\logq}{4\epsir}+\frac{1}{8}\ln^2{\frac{\mu^2}{-t}}+\frac{1}{2}\logq-\frac{\pi^2}{48}+1\right]\,.
\end{equation}
Its soft-collinear zero bin $I_\text{toy}^{[S]}$ results from power expanding the integrand using $k^\mu \sim Q\lambda$. This gives
\begin{align}
    I_\text{toy}^{[S]} &=\tilde{\mu}^{2\epsilon}\int\dlipsred{k}{d}\frac{(n\cdot k)\ (\nb\cdot p_1)}{[k^2+i0^+][(n\cdot k)(\nb\cdot p_1)+i0^+][(k+q_\perp)^2+i0^+]}\nonumber \\
    &=\tilde{\mu}^{2\epsilon}\int\dlipsred{k}{d}\frac{1}{[k^2+i0^+][(k+q_\perp)^2+i0^+]}\,,
\end{align}
which can be evaluated to obtain
\begin{equation}
    I_\text{toy}^{[S]}=\frac{i}{16\pi^2}\left[\frac{1}{\epsuv}+\logq+2\right]\,.
\end{equation}
This is an example of the soft-collinear zero bin \emph{not} being scaleless.

The Glauber and soft-Glauber-collinear zero bins for Eq.~\eqref{toy} are
\begin{align}
    &I_\text{toy}^{[G]}=\tilde{\mu}^{2\epsilon}\int\dlipsred{k}{d}\frac{(n\cdot k)\ (\nb\cdot p_1)}{[\vp{k}^2+i0^+][n\cdot (k+p_1)(\nb\cdot p_1)-(\vp{k}+\vec{p}_{1\perp})^2+i0^+][(\vp{k}+\vp{q})^2+i0^+]}\sim \cO(\lambda^2), \\*
    &I_\text{toy}^{[S][G]}=\tilde{\mu}^{2\epsilon}\int\dlipsred{k}{d}\frac{(n\cdot k)\ (\nb\cdot p_1)}{[\vp{k}^2+i0^+][(n\cdot k)(\nb\cdot p_1)+i0^+][(\vp{k}+\vp{q})^2+i0^+]}\sim \mathcal{O}(\lambda^2) \,.
\end{align}
Both of these zero bins are power suppressed relative to $\tilde{I}_\text{toy}$ and thus can be ignored at leading power. Thus, we find that the zero-bin subtracted toy integral is
\begin{align}
    I_\text{toy}&=\tilde{I}_\text{toy}-I_\text{toy}^{[S]}\\
    &=\frac{i}{16\pi^2}\left[\frac{1}{4\epsir^2}-\frac{1}{\epsuv}+\frac{2+\logq}{4\epsir}+\frac{1}{8}\ln^2{\frac{\mu^2}{-t}}-\frac{1}{2}\logq-\frac{\pi^2}{48}-1\right] \,.
\end{align}
Thus, the toy example shown here is an instance of a non-trivial zero bin subtraction that is essential to correctly match Glauber-SCET onto QCD. 

\subsection{Soft-Glauber zero bin \label{app:flowerzero}}

In this section we review the argument from~\cite{RS} about the zero-bin and regularization of the eikonal propagators.
For the sake of concreteness, we illustrate this using the soft flower graph  in Eq.~\eqref{Iflower}.
Stripping away the prefactors, the soft flower graph reduces to
\begin{align}
    I=\tilde{\mu}^{2\epsilon}\int \dlipsred{k}{d} \frac{1}{(\nb\cdt k \pm i0^+)(n\cdt k\pm i0^+)(k^2+i0^+)}\,.
\end{align}
The $\pm i 0^+$ refers to the choice for the poles of the eikonal propagators, depending on the choice of $S$
or $\overline S$ for the soft Wilson lines.

If we choose both sides to be the same, $+i0^+$, (i.e. pick $S_n$ and $S_{\nb}$), the integral evalutes to
\begin{equation}
    I_{++}=\frac{i}{16\pi^2}\left[-\frac{2}{\eta}\left(\frac{1}{\epsuv}-\frac{1}{\epsir}\right)+\ln(\frac{\mu^2}{\nu^2})\left(\frac{1}{\epsuv}-\frac{1}{\epsir}\right)+\frac{1}{\epsuv^2}-\frac{1}{\epsir^2}\right]\,,
\end{equation}
and the the soft-Glauber zero bin is 
\begin{equation}
    I^{[G]}_{++}=\tilde{\mu}^{2\epsilon}\int \dlipsred{k}{d} \frac{1}{(\nb\cdt k + i0^+)(n\cdt k + i0^+)(k_\perp^2+i0^+)}=0\,.
\end{equation}
The last line follows from the fact that the poles in $k^0$ are on the same side of the integration contour. Thus, for this choice of the signs of $i0^+$ in the eikonal denominators, the soft Glauber zero-bin explicitly vanishes. 

If one chooses opposite signs (i.e. pick $S_n$ and $\overline S_{\nb}$)
\begin{equation}
    I_{+-}=\tilde{\mu}^{2\epsilon}\int \dlipsred{k}{d} \frac{1}{(\nb\cdt k - i0^+)(n\cdt k + i0^+)(k^2+i0^+)}\ ,
\end{equation}
and the integral is different:
\begin{equation}
     I_{+-}= I_{++}+i \int\dlipsred{\vp{k}}{d-2} \int_{-\infty}^\infty \frac{dk_z}{2\pi}\nureg{|2k_z|}\frac{1}{(2k_z+i0^+)\ (\vp{k}^2-i0^+)}\,.
\end{equation}
The Glauber-soft zero bin in this case does not vanish:
\begin{align}
    I^{[G]}_{+-}&=\tilde{\mu}^{2\epsilon}\int \dlipsred{k}{d} \frac{1}{(\nb\cdt k - i0^+)(n\cdt k + i0^+)(k_\perp^2+i0^+)}\\
    &=i \int\dlipsred{\vp{k}}{d-2} \int_{-\infty}^\infty \frac{dk_z}{2\pi}\nureg{|2k_z|}\frac{1}{(2k_z+i0^+)\ (\vp{k}^2-i0^+)}\,,
\end{align}
where the last line follows from contour integration in $k^0$. Thus, we see that 
\begin{equation}
    I_{+ +} - I_{+ +}^{[G]}=I_{+ -} - I_{+ -}^{[G]}\,,
\end{equation}
and the zero-bin subtracted integrals are the same.
The soft-Glauber zero-bins allow one to be agnostic about the signs of the $i0^+$ in the eikonal denominators. The simplest choice is to have all the signs $+i0^+$ so that the soft-Glauber zero-bins vanish. 

%% file: integrals.tex
\section{Useful Loop Integrals}

In this appendix, we discuss how some of the rapidity divergent integrals can be computed, and give an example of an exact result with the soft eye graph.
\subsection{General formulas}
Rapidity divergent integrals from the collinear Wilson line graphs generally take the form 
\begin{align}
    \cI_W = \int \dlipsred{k}{d} \nureg{\nb\cdt k} \frac{f(\nb\cdt k ,\vp{k})}{[k^2+i0^+][(p+k)^2+i0^+]} \end{align}
where the numerator $f(\nb\cdt k, \vp{k})$ depends only on $\nb\cdot k$ and $\vp{k}$, and is independent of $n \cdot k$.
We first do the $k^+ = n\cdt k$ integral by contours. The poles in the $k^+$ plane are at
\begin{equation}
    k^+= \frac{\vp{k}^2-i0^+}{\nb\cdt k}\ , \quad \text{and}\quad  
    k^+= -n \cdot p + \frac{(\vp{k}+\vp{p})^2-i0^+}{\nb\cdt (p+k)}\ .
\end{equation}
Only when  $-\nb\cdt p<\nb\cdt k<0$ do the poles lie on opposite sides of the integration contour. If we close the contour in the upper half plane, we pick up the first pole. Changing variables to $z=-(\nb \cdot k)/(\nb \cdot p)$ gives
\begin{align}
      \cI_W 
    & =\frac{i}{4\pi} \left(\frac{\nu}{\nb \cdot p}\right)^\eta \int_{0 }^{1}{\rm d} z\ \frac{1}{z^\eta} \int\dlipsred{\vp{k}}{d-2} 
     \frac{f(-z \nb \cdot p, \vp{k})}{
\vp{k}^2+2 \vp{k} \cdot \vp{p} z +\vp{p}^2 z -p^+ p^- z(1-z) }   \nonumber \\
     & =\frac{i}{4\pi} \left(\frac{\nu}{\nb \cdot p}\right)^\eta \int_{0 }^{1}{\rm d} z\ \frac{1}{z^\eta} \int\dlipsred{\vp{k}}{d-2} \frac{f(-z \nb \cdot p, \vp{k})}{\vp{k}^2 + 2 \vp{k} \cdot \vp{p} z + \vp{p}^{\,2} z^2 }  \,,
\end{align}
using the on-shell condition $p^2=0$.
Shifting the integration momentum gives
\begin{align}
     \cI_W 
    & =\frac{i}{4\pi} \left(\frac{\nu}{\nb \cdot p}\right)^\eta \int_{0 }^{1}{\rm d} z\ \frac{1}{z^\eta} \int\dlipsred{\vp{k}}{d-2} \frac{f(-z \nb \cdot p, \vp{k}-z \vp{p})}{\vp{k}^2 }  \,.
    \label{eq:w}
\end{align}
The remaining computation, performing the $k_\perp$ integration and then the $z$ integration, depends on the particular integrand, but is generally straightforward. The result is proportional to $1/\epsuv-1/\epsir$.

Diagrams with triangle topology such as the collinear V graphs give rise to integrals of the form 
\begin{align}
    \cI_V = \int \dlipsred{k}{d} \nureg{\nb\cdt k} \frac{f(\nb\cdt k ,\vp{k})}{[k^2+i0^+][(p+k)^2+i0^+][(k+q_\perp)^2+i0^+]}\,.
\end{align}
There are now three poles in $n\cdt k$ given by
\begin{align}
    k^+ &= \frac{\vp{k}^2-i0^+}{\nb\cdt k}\ , \\ 
    k^+ &= -n\cdt p +\frac{\left(\vp{k}+\vp{p}\right)^2-i0^+}{\nb\cdt (p+k)}\ , \\
    k^+ &=\frac{(\vp{k}+\vp{q})^2-i0^+}{\nb\cdt k}\ .
\end{align}
Two of these poles are on opposite sides of the integration contour from the other only for the region $-\nb\cdt p<\nb\cdt k<0$. One can then close the contour in the lower half plane picking up the second pole, and change variables to $z=-(\nb \cdot k)/(\nb \cdot p)$ giving
\begin{align}
      \cI_V 
    & =-\frac{i}{4\pi} \left(\frac{\nu}{\nb \cdot p}\right)^\eta \int_{0 }^{1}{\rm d} z\ \frac{1-z}{z^\eta} \int\dlipsred{\vp{k}}{d-2} 
     \frac{f(-z \nb \cdot p, \vp{k})}{
\vp{k}^2+2 \vp{k} \cdot \vp{p} z + \vp{p} z -p^+ p^- z(1-z) } \times \nonumber \\
& \frac{1}{ (\vp{k} + \vp{q})^2 (1-z) +(\vp{k}+\vp{p})^2 z + \vp{p} z -p^+ p^- z(1-z) }  \,.
\end{align}
Combining denominators using Feynman parameters, using the on-shell condition $p^2=0$, and shifting the integration momentum
\begin{align}
      \cI_V 
    & =-\frac{i}{4\pi} \left(\frac{\nu}{\nb \cdot p}\right)^\eta \int_0^1 {\rm d} x \int_{0 }^{1}{\rm d} z\ \frac{1-z}{z^\eta} \int\dlipsred{\vp{k}}{d-2}  \times \nonumber \\
 &    \frac{f(-z \nb \cdot p, \vp{k}-z \vp{p}- x (1-z) \vp{q})}{
\left[ \vp{k}^2 + \vp{q}^2 x(1-x)(1-z)^2 - (2 \vp{p}- \vp{q}) \cdot \vp{q} x z(1-z)\right]^2 }  
  \,.
  \label{eq:v}
\end{align}
Finally, with our choice of kinematics, $2 \vp{p}=\vp{q}$, and the last term in the denominator vanishes.
Thus, \eqref{eq:v} can be used to evaluate rapidity divergent loop integrals coming from V graphs. All rapidity divergent integrals in this paper can be performed using \eqref{eq:v} and \eqref{eq:w}. In more generality the numerator could depend also on $n\cdot k = k^+$. In that case one simply substitutes the value of $k^+$ at the pole at which the residue is evaluated, after confirming that the integral is convergent as $k^+\to \infty$.

Using these formulas, we can compute the result of various representative terms in the $V$-graphs. \\
With a $k_\perp^2$ numerator ($Q = \nb \cdot p_1$):
\begin{multline}
   \tilde{\mu}^{2\epsilon}\int\dlipsred{k}{d} \frac{\nb\cdt p_1}{\nb\cdt k}\frac{k_\perp^2}{[k^2+i0^+][(k+p_1)^2+i0^+][(k+q_\perp)^2+i0^+]}\\
     =\frac{i}{16\pi^2}\bigg[\frac{1}{\eta}\left(\frac{1}{\epsuv}-\frac{1}{\epsir}\right)+\frac{1}{\epsuv}\left(1+\logp\right)-\frac{1}{4\epsir^2}-\frac{1}{\epsir}\left(\frac{1}{2}+\frac{1}{4}\logq+\logp\right) \\
     \hspace{3cm} -\frac{1}{8}\ln^2{\frac{\mu^2}{-t}}+\frac{1}{2}\logq+\frac{\pi^2}{48}+1\bigg]\,,
\end{multline}
With a constant numerator:
\begin{multline}
     \tilde{\mu}^{2\epsilon}\int\dlipsred{k}{d} \frac{\nb\cdt p_1}{\nb\cdt k}\frac{q_\perp^2}{[k^2+i0^+][(k+p_1)^2+i0^+][(k+q_\perp)^2+i0^+]}\\
    =\frac{i}{16\pi^2}\bigg[-\frac{2}{\eta}\left(\frac{1}{\epsir}+\logq\right)-\frac{1}{\epsir^2}-\frac{1}{\epsir}\left(\logq+2\logp\right)\\
    \hspace{3cm} -\frac{1}{2}\ln^2{\frac{\mu^2}{-t}}-2\logq\logp+\frac{3\pi^2}{4}\bigg] \,.
\end{multline}
With a numerator linear in $k_\perp$
\begin{multline}
     \tilde{\mu}^{2\epsilon}\int\dlipsred{k}{d} \frac{\nb\cdt p_1}{\nb\cdt k}\frac{k_\perp \cdt q_\perp}{[k^2+i0^+][(k+p_1)^2+i0^+][(k+q_\perp)^2+i0^+]}\\
    =\frac{i}{16\pi^2}\bigg[\frac{1}{\eta}\left(\frac{1}{\epsir}+\logq\right)+\frac{1}{2\epsir^2}+\frac{1}{\epsir}\left(\logp+\frac{1}{2}\logq\right)\\*
    +\frac{1}{4}\ln^2{\frac{\mu^2}{-t}}+\logq\logp -\frac{3\pi^2}{8}\bigg]\,.
\end{multline}

\newpage

\subsection{Eye Graph}\label{sec:eye}
As an example, we give the complete analytic result for the soft eye graph Eq.~\eqref{eq:eye}. Before expanding in $\eta$ or $\varepsilon$ we find
\begin{multline}
    \cI_{\text{eye}} 
    = \frac{2 g^4}{16\pi^2} T^a T^b C_F\cM_{T} \left(\frac{\mu^2 e^{\gamma_E}}{-t}\right)^{\varepsilon } 2^{2 \varepsilon } \pi \bigg[\frac{  \Gamma \left(\frac{1}{2}-\frac{\eta }{2}\right) (4-\eta -8 \varepsilon ) 
   \csc \left(\frac{\pi}{2}   (\eta +2 \varepsilon )\right)}{2\eta(\eta +2)\ \Gamma \left(\frac{3}{2}-\varepsilon -\frac{\eta }{2}\right)}\left(\frac{\nu ^2}{-t}\right)^{\eta /2} \\
   +\frac{\pi ^{1/2}  (2-3 \varepsilon) \csc (\pi  \varepsilon ) }{4\Gamma \left(\frac{3}{2}-\varepsilon \right)}\bigg]\,,
 \end{multline}
Note that the $\eta$ regulator is not needed for the 
 first term of the integrand, but included only for completeness.
 Expanding in $\eta$ to order $\eta^0$ gives
 \begin{multline}
  \cI_{\text{eye}}  =\frac{2 g^4}{16\pi^2} T^a T^b C_F \cM_{T} \left(\frac{\mu^2 e^{\gamma_E}}{-t}\right)^{\varepsilon } 2^{2 \varepsilon }\pi^{3/2} \csc (\pi \varepsilon) \bigg\{ \frac{1}{\eta} \frac{2 }{\Gamma\left(\frac12-\varepsilon \right)}  \\
 + \frac{ 1 }{4 \Gamma\left(\frac32-\varepsilon\right)} 
  \biggl[-1+\varepsilon + (2-4\varepsilon) \left(-\pi\cot (\pi \varepsilon) +  H_{\frac12-\varepsilon}
  + \ln \frac{4\nu^2}{-t} \right) \biggr]
   \bigg\} 
   \label{d2}
\end{multline}
where $H_n$ is the harmonic number. Expanding Eq.~\eqref{d2} in $\varepsilon$ gives 
\begin{multline}
    \cI_{\text{eye}}
    = \frac{2 g^4}{16\pi^2} T^a T^b C_F \cM_{T} \bigg\{ \frac{1}{\eta}
    \left[\frac{2}{\epsuv}+2\logq + \varepsilon\left(\ln^2 \frac{\mu^2}{-t}-\frac{\pi^2}{6} \right)\right]-\frac{1}{\epsuv^2}  + \frac{1}{\epsuv}\left(
    -\ln \frac{\mu^2}{\nu^2} 
    +\frac{3}{2}\right)  \\
     -\frac{1}{2}\ln^2 \frac{\mu^2}{-t}+\logq\ln{\frac{\nu^2}{-t}}  +\frac{3}{2}\logq-\frac{\pi^2}{12}+\frac{7}{2} + \varepsilon \biggl( -\frac16 \ln^3 \frac{\mu^2}{-t}  + \frac12 \ln^2 \frac{\mu^2}{-t} \ln \frac{\nu^2}{-t}   \\
   + \frac34 \ln^2 \frac{\mu^2}{-t}  - \frac{\pi^2}{12} \ln \frac{\mu^2}{-t}  +\frac72 \ln \frac{\mu^2}{-t} - \frac{\pi^2}{12} \ln \frac{\nu^2}{-t}  - \frac{\pi^2}{8} - \frac{14}{3} \zeta_3 + 7  \biggr) +\ldots \bigg\} \,,
\end{multline}
keeping terms to order $\varepsilon$.

%% file: main.bbl
\providecommand{\href}[2]{#2}\begingroup\raggedright\begin{thebibliography}{10}

\bibitem{Gell-Mann:1964aya}
M.~Gell-Mann, M.~Goldberger, F.~Low, E.~Marx, and F.~Zachariasen, {\it
  {Elementary Particles of Conventional Field Theory as Regge Poles. III}},
  {\em Phys. Rev.} {\bf 133} (1964) B145--B160.

\bibitem{Mandelstam:1965zz}
S.~Mandelstam, {\it {Non-Regge Terms in the Vector-Spinor Theory}},  {\em Phys.
  Rev.} {\bf 137} (1965) B949--B954.

\bibitem{McCoy:1976ff}
B.~M. McCoy and T.~T. Wu, {\it {Theory of Fermion Exchange in Massive Quantum
  Electrodynamics at High-Energy. 1.}},  {\em Phys. Rev. D} {\bf 13} (1976)
  369--378.

\bibitem{Grisaru:1973wbb}
M.~T. Grisaru, H.~J. Schnitzer, and H.-S. Tsao, {\it {Reggeization of
  elementary particles in renormalizable gauge theories --- vectors and
  spinors}},  {\em Phys. Rev. D} {\bf 8} (1973) 4498--4509.

\bibitem{Fadin:1975cb}
V.~S. Fadin, E.~A. Kuraev, and L.~N. Lipatov, {\it {On the Pomeranchuk
  Singularity in Asymptotically Free Theories}},  {\em Phys. Lett. B} {\bf 60}
  (1975) 50--52.

\bibitem{Lipatov:1976zz}
L.~N. Lipatov, {\it {Reggeization of the Vector Meson and the Vacuum
  Singularity in Nonabelian Gauge Theories}},  {\em Sov. J. Nucl. Phys.} {\bf
  23} (1976) 338--345.

\bibitem{Lipatov:1985uk}
L.~N. Lipatov, {\it {The Bare Pomeron in Quantum Chromodynamics}},  {\em Sov.
  Phys. JETP} {\bf 63} (1986) 904--912.

\bibitem{Pancheri:tot}
G.~Pancheri and Y.~N. Srivastava, {\it {Introduction to the physics of the
  total cross-section at LHC}: {A Review of Data and Models}},  {\em Eur. Phys.
  J. C} {\bf 77} (2017), no.~3 150,
  [\href{http://arxiv.org/abs/1610.10038}{{\tt arXiv:1610.10038}}].

\bibitem{Lipatov:1993yb}
L.~N. Lipatov, {\it {Asymptotic behavior of multicolor QCD at high energies in
  connection with exactly solvable spin models}},  {\em JETP Lett.} {\bf 59}
  (1994) 596--599, [\href{http://arxiv.org/abs/hep-th/9311037}{{\tt
  hep-th/9311037}}].

\bibitem{Faddeev:1994zg}
L.~D. Faddeev and G.~P. Korchemsky, {\it {High-energy QCD as a completely
  integrable model}},  {\em Phys. Lett. B} {\bf 342} (1995) 311--322,
  [\href{http://arxiv.org/abs/hep-th/9404173}{{\tt hep-th/9404173}}].

\bibitem{DelDuca:2001gu}
V.~Del~Duca and E.~W.~N. Glover, {\it {The High-energy limit of QCD at two
  loops}},  {\em JHEP} {\bf 10} (2001) 035,
  [\href{http://arxiv.org/abs/hep-ph/0109028}{{\tt hep-ph/0109028}}].

\bibitem{DelDuca:2013ara}
V.~Del~Duca, G.~Falcioni, L.~Magnea, and L.~Vernazza, {\it {High-energy QCD
  amplitudes at two loops and beyond}},  {\em Phys. Lett. B} {\bf 732} (2014)
  233--240, [\href{http://arxiv.org/abs/1311.0304}{{\tt arXiv:1311.0304}}].

\bibitem{DelDuca:2014cya}
V.~Del~Duca, G.~Falcioni, L.~Magnea, and L.~Vernazza, {\it {Analyzing
  high-energy factorization beyond next-to-leading logarithmic accuracy}},
  {\em JHEP} {\bf 02} (2015) 029, [\href{http://arxiv.org/abs/1409.8330}{{\tt
  arXiv:1409.8330}}].

\bibitem{DelDuca:2011ae}
V.~Del~Duca, C.~Duhr, E.~Gardi, L.~Magnea, and C.~D. White, {\it {The Infrared
  structure of gauge theory amplitudes in the high-energy limit}},  {\em JHEP}
  {\bf 12} (2011) 021, [\href{http://arxiv.org/abs/1109.3581}{{\tt
  arXiv:1109.3581}}].

\bibitem{Caron-Huot:2019vjl}
S.~Caron-Huot, L.~J. Dixon, F.~Dulat, M.~von Hippel, A.~J. McLeod, and
  G.~Papathanasiou, {\it {Six-Gluon amplitudes in planar $ \mathcal{N} $ = 4
  super-Yang-Mills theory at six and seven loops}},  {\em JHEP} {\bf 08} (2019)
  016, [\href{http://arxiv.org/abs/1903.10890}{{\tt arXiv:1903.10890}}].

\bibitem{DelDuca:2018rhj}
V.~Del~Duca, S.~Druc, J.~Drummond, C.~Duhr, F.~Dulat, R.~Marzucca,
  G.~Papathanasiou, and B.~Verbeek, {\it {Amplitudes in the Multi-Regge Limit
  of $\mathcal{N}$=4 SYM}},  {\em Acta Phys. Polon. Supp.} {\bf 12} (2019),
  no.~4 961--966, [\href{http://arxiv.org/abs/1811.10588}{{\tt
  arXiv:1811.10588}}].

\bibitem{DelDuca:2016lad}
V.~Del~Duca, S.~Druc, J.~Drummond, C.~Duhr, F.~Dulat, R.~Marzucca,
  G.~Papathanasiou, and B.~Verbeek, {\it {Multi-Regge kinematics and the moduli
  space of Riemann spheres with marked points}},  {\em JHEP} {\bf 08} (2016)
  152, [\href{http://arxiv.org/abs/1606.08807}{{\tt arXiv:1606.08807}}].

\bibitem{Basso:2014pla}
B.~Basso, S.~Caron-Huot, and A.~Sever, {\it {Adjoint BFKL at finite coupling: a
  short-cut from the collinear limit}},  {\em JHEP} {\bf 01} (2015) 027,
  [\href{http://arxiv.org/abs/1407.3766}{{\tt arXiv:1407.3766}}].

\bibitem{Dixon:2011pw}
L.~J. Dixon, J.~M. Drummond, and J.~M. Henn, {\it {Bootstrapping the three-loop
  hexagon}},  {\em JHEP} {\bf 11} (2011) 023,
  [\href{http://arxiv.org/abs/1108.4461}{{\tt arXiv:1108.4461}}].

\bibitem{Bartels:2014mka}
J.~Bartels, V.~Schomerus, and M.~Sprenger, {\it {The Bethe roots of Regge cuts
  in strongly coupled $ \mathcal{N}=4 $ SYM theory}},  {\em JHEP} {\bf 07}
  (2015) 098, [\href{http://arxiv.org/abs/1411.2594}{{\tt arXiv:1411.2594}}].

\bibitem{Kuraev:1976ge}
E.~A. Kuraev, L.~N. Lipatov, and V.~S. Fadin, {\it {Multi-Reggeon Processes in
  the Yang-Mills Theory}},  {\em Sov. Phys. JETP} {\bf 44} (1976) 443--450.

\bibitem{Balitsky:1978ic}
I.~I. Balitsky and L.~N. Lipatov, {\it {The Pomeranchuk Singularity in Quantum
  Chromodynamics}},  {\em Sov. J. Nucl. Phys.} {\bf 28} (1978) 822--829.

\bibitem{kovchegov}
Y.~V. Kovchegov and E.~Levin, {\em {Quantum chromodynamics at high energy}},
  vol.~33.
\newblock Cambridge University Press, 8, 2012.

\bibitem{forshaw}
J.~R. Forshaw and D.~A. Ross, {\em {Quantum chromodynamics and the pomeron}},
  vol.~9.
\newblock Cambridge University Press, 1, 2011.

\bibitem{bauerBXs}
C.~W. Bauer, S.~Fleming, and M.~E. Luke, {\it {Summing Sudakov logarithms in $B
  \to X_s \gamma$ in effective field theory}},  {\em Phys. Rev. D} {\bf 63}
  (2000) 014006, [\href{http://arxiv.org/abs/hep-ph/0005275}{{\tt
  hep-ph/0005275}}].

\bibitem{bauer2001effective}
C.~W. Bauer, S.~Fleming, D.~Pirjol, and I.~W. Stewart, {\it {An Effective field
  theory for collinear and soft gluons: Heavy to light decays}},  {\em Phys.
  Rev. D} {\bf 63} (2001) 114020,
  [\href{http://arxiv.org/abs/hep-ph/0011336}{{\tt hep-ph/0011336}}].

\bibitem{bauer2001invariant}
C.~W. Bauer and I.~W. Stewart, {\it {Invariant operators in collinear effective
  theory}},  {\em Phys. Lett. B} {\bf 516} (2001) 134--142,
  [\href{http://arxiv.org/abs/hep-ph/0107001}{{\tt hep-ph/0107001}}].

\bibitem{bauer2002hard}
C.~W. Bauer, S.~Fleming, D.~Pirjol, I.~Z. Rothstein, and I.~W. Stewart, {\it
  {Hard scattering factorization from effective field theory}},  {\em Phys.
  Rev. D} {\bf 66} (2002) 014017,
  [\href{http://arxiv.org/abs/hep-ph/0202088}{{\tt hep-ph/0202088}}].

\bibitem{bauer2002power}
C.~W. Bauer, D.~Pirjol, and I.~W. Stewart, {\it {Power counting in the soft
  collinear effective theory}},  {\em Phys. Rev. D} {\bf 66} (2002) 054005,
  [\href{http://arxiv.org/abs/hep-ph/0205289}{{\tt hep-ph/0205289}}].

\bibitem{bauer2002soft}
C.~W. Bauer, D.~Pirjol, and I.~W. Stewart, {\it {Soft collinear factorization
  in effective field theory}},  {\em Phys. Rev. D} {\bf 65} (2002) 054022,
  [\href{http://arxiv.org/abs/hep-ph/0109045}{{\tt hep-ph/0109045}}].

\bibitem{RS}
I.~Z. Rothstein and I.~W. Stewart, {\it {An Effective Field Theory for Forward
  Scattering and Factorization Violation}},  {\em JHEP} {\bf 08} (2016) 025,
  [\href{http://arxiv.org/abs/1601.04695}{{\tt arXiv:1601.04695}}].

\bibitem{Frye:2018xjj}
C.~Frye, H.~Hannesdottir, N.~Paul, M.~D. Schwartz, and K.~Yan, {\it {Infrared
  Finiteness and Forward Scattering}},  {\em Phys. Rev. D} {\bf 99} (2019),
  no.~5 056015, [\href{http://arxiv.org/abs/1810.10022}{{\tt
  arXiv:1810.10022}}].

\bibitem{Hannesdottir:2019opa}
H.~Hannesdottir and M.~D. Schwartz, {\it {$S$ -Matrix for massless particles}},
   {\em Phys. Rev. D} {\bf 101} (2020), no.~10 105001,
  [\href{http://arxiv.org/abs/1911.06821}{{\tt arXiv:1911.06821}}].

\bibitem{Hannesdottir:2019rqq}
H.~Hannesdottir and M.~D. Schwartz, {\it {A Finite $S$-Matrix}},
  \href{http://arxiv.org/abs/1906.03271}{{\tt arXiv:1906.03271}}.

\bibitem{marcantonini2009reparametrization}
C.~Marcantonini and I.~W. Stewart, {\it {Reparameterization Invariant Collinear
  Operators}},  {\em Phys. Rev. D} {\bf 79} (2009) 065028,
  [\href{http://arxiv.org/abs/0809.1093}{{\tt arXiv:0809.1093}}].

\bibitem{Manohar:2006nz}
A.~V. Manohar and I.~W. Stewart, {\it {The Zero-Bin and Mode Factorization in
  Quantum Field Theory}},  {\em Phys. Rev. D} {\bf 76} (2007) 074002,
  [\href{http://arxiv.org/abs/hep-ph/0605001}{{\tt hep-ph/0605001}}].

\bibitem{MSSV}
I.~Moult, M.~P. Solon, I.~W. Stewart, and G.~Vita, {\it {Fermionic Glauber
  Operators and Quark Reggeization}},  {\em JHEP} {\bf 02} (2018) 134,
  [\href{http://arxiv.org/abs/1709.09174}{{\tt arXiv:1709.09174}}].

\bibitem{chiu2012formalism}
J.-Y. Chiu, A.~Jain, D.~Neill, and I.~Z. Rothstein, {\it {A Formalism for the
  Systematic Treatment of Rapidity Logarithms in Quantum Field Theory}},  {\em
  JHEP} {\bf 05} (2012) 084, [\href{http://arxiv.org/abs/1202.0814}{{\tt
  arXiv:1202.0814}}].

\bibitem{duffexponentialreg}
Y.~Li, D.~Neill, and H.~X. Zhu, {\it {An exponential regulator for rapidity
  divergences}},  {\em Nucl. Phys. B} {\bf 960} (2020) 115193,
  [\href{http://arxiv.org/abs/1604.00392}{{\tt arXiv:1604.00392}}].

\bibitem{chiudeltareg}
J.-y. Chiu, A.~Fuhrer, A.~H. Hoang, R.~Kelley, and A.~V. Manohar, {\it
  {Soft-Collinear Factorization and Zero-Bin Subtractions}},  {\em Phys. Rev.
  D} {\bf 79} (2009) 053007, [\href{http://arxiv.org/abs/0901.1332}{{\tt
  arXiv:0901.1332}}].

\bibitem{Chiu:2009mg}
J.-y. Chiu, A.~Fuhrer, R.~Kelley, and A.~V. Manohar, {\it {Factorization
  Structure of Gauge Theory Amplitudes and Application to Hard Scattering
  Processes at the LHC}},  {\em Phys. Rev. D} {\bf 80} (2009) 094013,
  [\href{http://arxiv.org/abs/0909.0012}{{\tt arXiv:0909.0012}}].

\bibitem{ChiuRRGE}
J.-y. Chiu, A.~Jain, D.~Neill, and I.~Z. Rothstein, {\it {The Rapidity
  Renormalization Group}},  {\em Phys. Rev. Lett.} {\bf 108} (2012) 151601,
  [\href{http://arxiv.org/abs/1104.0881}{{\tt arXiv:1104.0881}}].

\bibitem{packageX}
H.~H. Patel, {\it {Package-X 2.0: A Mathematica package for the analytic
  calculation of one-loop integrals}},  {\em Comput. Phys. Commun.} {\bf 218}
  (2017) 66--70, [\href{http://arxiv.org/abs/1612.00009}{{\tt
  arXiv:1612.00009}}].

\bibitem{Ellis:1985er}
R.~K. Ellis and J.~C. Sexton, {\it {QCD Radiative Corrections to Parton Parton
  Scattering}},  {\em Nucl. Phys. B} {\bf 269} (1986) 445--484.

\bibitem{Fuhrer:2010eu}
A.~Fuhrer, A.~V. Manohar, J.-y. Chiu, and R.~Kelley, {\it {Radiative
  Corrections to Longitudinal and Transverse Gauge Boson and Higgs
  Production}},  {\em Phys. Rev. D} {\bf 81} (2010) 093005,
  [\href{http://arxiv.org/abs/1003.0025}{{\tt arXiv:1003.0025}}].

\bibitem{Hahn:2000kx}
T.~Hahn, {\it {Generating Feynman diagrams and amplitudes with FeynArts 3}},
  {\em Comput. Phys. Commun.} {\bf 140} (2001) 418--431,
  [\href{http://arxiv.org/abs/hep-ph/0012260}{{\tt hep-ph/0012260}}].

\bibitem{Shtabovenko:2020gxv}
V.~Shtabovenko, R.~Mertig, and F.~Orellana, {\it {FeynCalc 9.3: New features
  and improvements}},  {\em Comput. Phys. Commun.} {\bf 256} (2020) 107478,
  [\href{http://arxiv.org/abs/2001.04407}{{\tt arXiv:2001.04407}}].

\bibitem{Shtabovenko:2016whf}
V.~Shtabovenko, {\it {FeynHelpers: Connecting FeynCalc to FIRE and Package-X}},
   {\em Comput. Phys. Commun.} {\bf 218} (2017) 48--65,
  [\href{http://arxiv.org/abs/1611.06793}{{\tt arXiv:1611.06793}}].

\bibitem{mattcollinearviolation}
M.~D. Schwartz, K.~Yan, and H.~X. Zhu, {\it {Collinear factorization violation
  and effective field theory}},  {\em Phys. Rev. D} {\bf 96} (2017), no.~5
  056005, [\href{http://arxiv.org/abs/1703.08572}{{\tt arXiv:1703.08572}}].

\bibitem{css}
C.~F. Berger, T.~Kucs, and G.~F. Sterman, {\it {Event shape / energy flow
  correlations}},  {\em Phys. Rev. D} {\bf 68} (2003) 014012,
  [\href{http://arxiv.org/abs/hep-ph/0303051}{{\tt hep-ph/0303051}}].

\bibitem{Feige:2014wja}
I.~Feige and M.~D. Schwartz, {\it {Hard-Soft-Collinear Factorization to All
  Orders}},  {\em Phys. Rev. D} {\bf 90} (2014), no.~10 105020,
  [\href{http://arxiv.org/abs/1403.6472}{{\tt arXiv:1403.6472}}].

\bibitem{regge}
I.~Moult, S.~Raman, G.~Ridgway, and I.~Stewart. in preparation.

\end{thebibliography}\endgroup
